\documentclass[prd,twocolumn,showpacs,nofootinbib,superscriptaddress]{revtex4}

\usepackage{amsfonts}
\usepackage{amsmath}
\usepackage{amssymb}
\usepackage{upgreek}
\usepackage{bm}
\usepackage{dcolumn}
\usepackage{epsfig}
\usepackage{graphicx}
\usepackage{graphics}
\usepackage{latexsym}
\usepackage{rotating}
\usepackage{hyperref}
\usepackage{mathtools}
\usepackage{xspace} % Sensible space treatment at end of simple macros
\usepackage[usenames,dvipsnames]{color}
\usepackage[latin1]{inputenc}

\usepackage{ulem}
\definecolor {darkgreen}{rgb}{0.2,0.7,0.2} 
\allowdisplaybreaks

\DeclareSymbolFont{matha}{OML}{txmi}{m}{it}% txfonts

\newcommand\be{\begin{equation}}
\newcommand\ba{\begin{eqnarray}}
\newcommand\ee{\end{equation}}
\newcommand\ea{\end{eqnarray}}

\newcommand{\mb}[1]{\mbox{\boldmath $#1$}}

\newcommand{\nn}{\nonumber}

\newcommand{\av}{{\mbox{\tiny av}}}

\newcommand{\pr}{{\mbox{\tiny pr}}}
\newcommand{\rr}{{\mbox{\tiny rr}}}

\newcommand{\beq}{\begin{equation}}
\newcommand{\eeq}{\end{equation}}
\newcommand{\bes}{\begin{subequations}}
\newcommand{\ees}{\end{subequations}}
\newcommand{\beqn}{\begin{eqnarray*}}
\newcommand{\eeqn}{\end{eqnarray*}}      
\newcommand{\p}{\partial}
\newcommand{\dvec}[1]{\dot{\bm{#1}}}
\newcommand{\uvec}[1]{\bm{\hat{#1}}}
\newcommand{\duvec}[1]{\dot{\bm{\hat{#1}}}}
\newcommand{\sub}[1]{_{\text{#1}}}

\newcommand{\ord}[1]{\mathcal{O} \left( #1 \right)}
\newcommand{\precav}[1]{\left\langle #1 \right\rangle\sub{\pr}}
\newcommand{\precphi}[1]{\left\langle #1 \right\rangle\sub{$\phi_z$}}

\def\nn{\nonumber}

\begin{document}

\title{Constructing gravitational waves from generic spin-precessing compact binary inspirals}

\author{Katerina Chatziioannou}
\affiliation{eXtreme Gravity Institute, Department of Physics, Montana State University, Bozeman, Montana 59717, USA}
\affiliation{Canadian Institute for Theoretical Astrophysics, 60 St. George Street, University of Toronto, Toronto, ON M5S 3H8, Canada}
\author{Antoine Klein}
\affiliation{Department of Physics and Astronomy, The University of Mississippi, University, MS 38677, USA}
\affiliation{CENTRA, Departamento de F\'isica, Instituto Superior T\'ecnico, 
Universidade de Lisboa, Avenida Rovisco Pais 1, 1049 Lisboa, Portugal}
\affiliation{CNRS, UMR 7095, Institut d'Astrophysique de Paris, 98 bis Bd 
Arago, 
75014 Paris, France}
\author{Nicol\'as Yunes}
\affiliation{eXtreme Gravity Institute, Department of Physics, Montana State University, Bozeman, Montana 59717, USA}
\author{Neil Cornish}
\affiliation{eXtreme Gravity Institute, Department of Physics, Montana State University, Bozeman, Montana 59717, USA}

\date{\today}

%%%%%%%%%%%%%%%%%%%%%%%%%%%%%%%%%%%%%%%%%%%%%%%
\begin{abstract}

The coalescence of compact objects is one of the most promising sources, as well as the source of the first detections, of gravitational waves for ground-based interferometric detectors, such as advanced LIGO and Virgo. 
Generically, compact objects in binaries are expected to be spinning with spin angular momenta misaligned with the orbital angular momentum, causing the orbital plane to precess. 
This precession adds rich structure to the gravitational waves, introducing such complexity that an analytic closed-form description has been unavailable until now. 
We here construct the first closed-form frequency-domain gravitational waveforms that are valid for generic spin-precessing quasicircular compact binary inspirals.
We first construct time-domain gravitational waves by solving the post-Newtonian precession equations of motion with radiation reaction through multiple scale analysis. 
We then Fourier transform these time-domain waveforms with the method of shifted uniform asymptotics to obtain closed-form expressions for frequency-domain waveforms.
We study the accuracy of these analytic, frequency-domain waveforms relative to waveforms obtained by numerically evolving the post-Newtonian equations of motion and find that they are suitable for unbiased parameter estimation for $99.2\%(94.6\%)$ of the binary configurations we studied at a signal-to-noise ratio of $10(25)$.
These new frequency-domain waveforms could be used for detection and parameter estimation studies due to their accuracy and low computational cost.

\end{abstract}
%%%%%%%%%%%%%%%%%%%%%%%%%%%%%%%%%%%%%%%%%%%%%%%
\pacs{04.80.Nn,04.30.-w,97.60.Jd}

\maketitle

%%%%%%%%%%%%%%%%%%%%%%%%%%%%%%%%%%%%%%%%%%%%%%%
\section{Introduction}
\label{intro}

%Spin is ubiquitous. Describe formation mechanism of BH binaries and misalignment of L with S 
Spin is ubiquitous in Nature; under the influence of a generic perturbation any 
astrophysical system will rotate even if the initial state was perfectly 
spherically symmetric. When massive stars give birth to neutron 
stars (NSs) or black holes (BHs) through supernova explosions, the newly-born remnant typically spins rapidly even if the progenitor was spinning 
slowly. This is possibly due to asymmetries in the 
supernova explosion inducing a ``kick'' on the remnant causing it to rotate~\cite{Spruit:1998sg}. At a fundamental level, the 
physics at play here is the same as that which causes 
a soccer ball to spin after kicked.

%Compact binaries, in particular, will thus be spinning with misaligned spins. 
The spin angular momenta of the components of a compact binary system will not necessarily be aligned with the orbital angular momentum. For example, consider an isolated binary system of two stars with spins aligned with the orbital angular momentum. The most massive star will first fill its Roche lobe and transfer mass to the companion before going supernova. The compact remnant (BH or NS) receives a kick spinning it up~\cite{Spruit:1998sg} and tilting the orbital plane, since the kick's direction is typically correlated with the spin of the exploding star~\cite{Wang:2005jg,Wang:2006zia}; the various angular momenta become misaligned~\cite{Kalogera:1999tq,Gerosa:2014kta}. Eventually, the second star also fills its Roche lobe and the binary enters a common envelope phase. In this phase, the angular momenta could partially align through tidal effects. But, again, the phase ends with the star going supernova and endowing the binary with a kick that spins the remnant up and typically tilts the orbital plane, misaligning the angular momenta yet again~\cite{Kalogera:1999tq,Gerosa:2014kta}. 

% Nothing to damp spin or to align them with L by the time they enter the LIGO band. 
Few mechanism exist that could prevent misalignment or realign the spins of compact binary components with the orbital angular momentum. One possibility is if the supernova kicks are in the orbital plane, i.e.~\emph{perpendicular} to the spin angular momentum, such that the orbital plane is not tilted~\cite{Kalogera:1999tq}. This possibility is remote, with models and data suggesting that the kick is actually \emph{aligned} with the spin angular momentum~\cite{Wang:2005jg,Wang:2006zia,Kaplan:2008qm}. A mechanism for realignment is through torques exerted by a circumbinary accretion disk on BH binaries~\cite{Bogdanovic:2007hp}. Such disks, however, are only expected in galaxy mergers of supermassive BHs, whose gravitational waves (GWs) would be outside of the sensitivity band of the ground-based detectors advanced LIGO (aLIGO) and advanced Virgo (AdV).  Besides the formation mechanism described above, dynamical formation channels are expected to result in binaries with arbitrarily distributed spin directions~\cite{Rodriguez:2016vmx}. Finally, we note that the binary might undergo spin-orbit resonances during its evolution~\cite{Schnittman:2004vq}. These resonances, however, do not align the spins with the orbital angular moment, but rather maintain certain special (``resonant'') precessing configurations~\cite{Schnittman:2004vq,Gerosa:2014kta,Kesden:2014sla,Gerosa:2015tea,Gerosa:2015hba}.

% How does spin affect the waves that LIGO will see. 
Compact binaries emitting GWs in the sensitivity band of aLIGO and AdV 
will thus have spins with arbitrary magnitudes and directions, though the spin of NSs will typically be smaller than that of BHs. 
Such GWs are very different from those emitted 
by nonspinning binaries or binaries with spins aligned with the orbital angular 
momentum for one main reason: \emph{precession}. When spins are misaligned with 
the orbital angular momentum, all angular momenta precess about the total angular 
momentum, causing the orbital plane to precess too. 
This induces amplitude and phase modulations in the GWs, for example through the changing inclination angle of the system relative to the line of sight. Accurately 
modeling these modulations can be important for detection and parameter estimation of GW sources~\cite{Chatziioannou:2014coa,Chatziioannou:2014bma}. 

%Signals LIGO will detect will be buried in the noise, so you have to model/filter data with templates. 
GWs with such rich and complex structure are double-edged swords: on the one 
hand, this structure can encode new information about the source and break degeneracies in parameter estimation; 
on the other hand, this intrinsic complexity comes at the cost of an 
increased difficulty to model these waves. First, the temporal evolution 
of the orbital phase depends on the 
angular momenta, which themselves satisfy certain precession equations, 
increasing the overall complexity of the differential system. Second, precession 
introduces \emph{mathematical catastrophes} when computing the Fourier transform 
of the GWs: degenerate critical points in the orbital phase, where the 
first and second time derivatives vanish, violating the assumptions of  the standard 
\emph{stationary phase approximation}\footnote{The SPA is the leading-order term in the 
asymptotic expansion of a Fourier integral through the method of steepest 
descent~\cite{Droz:1999qx,Bender}.} (SPA) and rendering it nonapplicable~\cite{Klein:2014bua}. 

%Modeling GWs is important for LIGO. Detection and PE depends on accuracy of templates.
Accurate modeling of GWs is important for detection and crucial for parameter estimation with ground-based detectors, since the expected signals might be deeply buried in the detector noise. The optimal strategy for extracting known signals from noise is by fitting \emph{waveform models} to the interferometric data and minimizing the residual. The efficiency of this method relies on the accuracy of the waveform model, with parameter estimation placing more stringent requirements on the accuracy of the models used. Inaccuracies in the models can lead to missed signals or systematic errors in the extracted parameters. 

%History of spin-precessing templates until now. 
This has motivated the construction of waveform models for coalescing compact binaries. During the inspiral, the binary can be modeled with the \emph{post-Newtonian} (PN) formalism, an expansion in small characteristic velocities and weak gravitational fields~\cite{Blanchet:2014}. When the binary components are not spinning or when their spin is aligned/anti-aligned with the orbital angular momentum, the equations of motion have been derived and solved up to 3.5PN order\footnote{A term is of $A$PN order is proportional to $(u/c)^{2A}$ relative to its controlling factor, where $u$ is some characteristic velocity and $c$ the speed of light.} including radiation reaction due to GW energy loss. When the spins are misaligned with the orbital angular momentum, the orbital equations of motion and the precession equations have been derived to 2.5PN order. In this case, a closed form solution has not been obtained due to the complexity of the differential system. 

%Apostolatos, EoB with Spin, Hannam
To this day, four main representations of GWs from spin-precessing compact binaries exist. The first representation is based on the fact that the precession equations admit a closed-form analytic solution when only one object is spinning~\cite{Apostolatos:1994mx,PhysRevD.52.605,PhysRevD.54.2421,PhysRevD.54.2438,Hannam:2013oca}. The ensuing motion is~\emph{simple precession} and the resulting waveform is ideal for BHNS systems~\cite{Lundgren:2013jla}. The second representation is based on the~\emph{effective-one-body} formulation of the general relativistic two-body problem~\cite{Buonanno:1998gg,Buonanno99,Damour01}. The resulting waveform is ever improving through fits of its nonprecessing part to numerical relativity simulations~\cite{Taracchini:2013rva,Pan:2013rra,PhysRevD.93.044007} and describes the full coalescence, albeit at the expense of prohibitive computational cost. The third representation utilizes a coordinate frame in which precessional effects are minimized~\cite{PhysRevD.84.124011,Schmidt:2010it} to compute a simpler waveform~\cite{PhysRevD.93.044007} and map it back to the source frame~\cite{Schmidt:2012rh,Schmidt:2014iyl}. This approach applies to the full binary coalescence and the waveform was found to be sufficiently good for detection, but could introduce biases in parameter estimation~\cite{Hannam:2013oca}\footnote{It is worth emphasizing, thought, that the first GW detection~\cite{TheLIGOScientific:2016wfe,Abbott:2016izl} did not suffer from such systematics due to its orientation and minimal precession~\cite{TheLIGOScientific:2016wfe,Abbott:2016wiq}.}.

%MSA
The final representation of GWs from inspiraling spin-precessing systems was through a multiple scale analysis (MSA)~\cite{Bender}, a well-known mathematical technique to solve differential systems that have distinct characteristic scales by expanding in the ratio of these scales. For the problem at hand, the orbital time scale is much shorter than the precession time scale, which in turn is much shorter than the radiation reaction time scale. This technique has already been applied successfully to nearly aligned~\cite{Klein:2013qda} and slowly spinning~\cite{Chatziioannou:2013dza} systems. The latter are accurate representations of NSNS inspirals, both for detection and parameter estimation~\cite{Chatziioannou:2014bma,Chatziioannou:2014coa,Chatziioannou:2015uea}. In this paper we utilize two recent breakthroughs to construct waveforms for spin-precessing systems with arbitrary spin magnitudes and orientations with MSA.

%Describe non-phenomenological: KESDEN
The first breakthrough in the modeling of generic spin-precessing binaries was by Kesden et al.~\cite{Kesden:2014sla,Gerosa:2015hba,Gerosa:2015tea}. Neglecting radiation reaction, the authors found an \emph{exact} solution to the precession equations that govern the evolution of the orbital and the spin angular momenta of the binary. By identifying certain constants of the precessional motion, they were able to express all angular momenta as functions of the total spin magnitude, which satisfies an ordinary differential equation. We here solve this differential equation analytically and obtain an exact solution to the precession equations in the absence of radiation reaction. We then use MSA to introduce radiation reaction perturbatively as an expansion in the ratio of the precession to the radiation reaction time scale. With this at hand, we obtain time-domain waveforms in terms of the parameters of the system only. 

%Describe non-phenomenological: SUA
The second breakthrough in the modeling of generic spin-precessing binaries was by Klein et al.~\cite{Klein:2014bua}, and tackles the failure of the SPA.  The authors introduced the 
\emph{shifted uniform asymptotics} (SUA) method where the waveform is decomposed into Bessel 
functions, the Fourier integral is evaluated term by term 
in the SPA, and then resumed using the exponential 
shift theorem. The result is a closed-form analytic expression for the 
gravitational wave in the frequency domain as a series of time-domain 
waveforms evaluated at shifted stationary times. Unlike previous approaches, both the time- and frequency-domain waveforms that we obtain are valid for arbitrary mass ratios, arbitrary spin magnitudes and arbitrary spin orientations. This waveform was first presented in Ref.~\cite{Chatziioannou:2016ezg}, while this paper provides the details of its derivation.

%Is this useful? State something about computational efficiency. 
Closed-form expressions for the waveforms have several advantages. From 
a theoretical standpoint, analytic solutions shed light on the physical 
processes at play, the structure of the resultant signal, and the transition through different 
resonant states. From a 
practical standpoint, analytic solutions are in general faster to evaluate, avoiding costly numerical integrations and discrete Fourier transforms. Estimating the computational gain from closed-form, analytic expressions relative to numerical 
ones is not straightforward since it depends heavily on the implementation. 
However, we estimate that in the restricted waveform case (when only one 
harmonic is used) the closed-form, analytic frequency-domain waveforms computed 
here can be an order of magnitude faster than the implementation 
of~\cite{Klein:2014bua}, and are at worst comparable.

\begin{table}
\begin{centering}
\begin{tabular}{ccccccccc}
\hline
\hline
\noalign{\smallskip}
{}   &&  \multicolumn{1}{c}{NSNS} &&  \multicolumn{1}{c}{BHNS} &&  \multicolumn{1}{c}{BHBH} &&  \multicolumn{1}{c}{HSNSBH}  \\
\hline
\noalign{\smallskip}
$m_1$                && 1.6$M_{\odot}$ && 10$M_{\odot}$   && 10$M_{\odot}$ && 10$M_{\odot}$ \\
$m_2$                && 1.4$M_{\odot}$ && 1.4$M_{\odot}$  && 5$M_{\odot}$ && 1.4$M_{\odot}$\\
$\cos{\theta_L}$ && 1    && 1     && 1 && 1\\
$\phi_L$             && 0    && 0     && 0 && 0\\
$\cos{\theta_1}$ && 0.5 && 0.5  && 0.5 && 0.5\\
$\phi_1$             && 1.2  && 1.2  && 1.2 && 1.2\\
$\chi_1$             && 0.08 && 0.7  && 0.7 && 0.7\\
$\cos{\theta_2}$ && 0.7  && 0.7  && 0.7 && 0.7\\
$\phi_2$             && 2.5  && 2.5  && 2.5 && 2.5\\
$\chi_2$             && 0.1  && 0.1   && 0.6 && 0.6\\
\noalign{\smallskip}
\hline
\hline
\end{tabular}
\end{centering}
\caption{Parameters of the systems we use for comparisons of our analytic solution to the numerical solution to the PN precession equations. All parameters are defined at $50$Hz and in a frame were the orbital angular momentum is aligned with the $z$ axis.}
\label{table:systems}
\end{table}

%Notation and Conventions. 
The remainder of the paper provides the details of the waveform construction described above. Throughout, we use geometric units where $G=c=1$ and use the following conventions:
\begin{itemize}
\item
Vectors are written in boldface, with components $\bm{A}=[A_x,A_y,A_z]$ and magnitude $A$. Unit vectors are denoted with a hat, e.g. $\bm{\hat{A}}$.
\item
The masses of the two binary components are $m_A$, with $A \in \{1,2\}$, 
the total mass $M\equiv m_1+m_2$ is set equal to 1, the mass ratio is 
$q\equiv m_2/m_1 < 1$, the symmetric mass ratio is $\eta\equiv m_1 m_2$ and the 
mass difference is $\delta m = m_{1} - m_{2}$. 
\item
The Newtonian orbital angular momentum of the system is $\bm{L}$, the spin angular momentum of each body is $\bm{S}_A$, and the total angular momentum is $\bm{J} = \bm{L} + \bm{S}_{1} + \bm{S}_{2}$. The dimensionless spin parameter of each object is $\chi_A\equiv S_A/m_A^2$ with $A \in \{1,2\}$. 
\item
The orbital angular frequency in a frame fixed to the orbital plane is $\omega$, while the PN expansion parameter we use is $ v\equiv \omega^{1/3}=  \eta L^{-1}$.
\item
We test our analytic solution by comparing it to the numerical solution to the precession equations for certain systems. We select a NSNS, a BHNS, a BHBH, and a highly spinning (HS)NSBH system with parameters given in Table~\ref{table:systems} in a frame were the $z$ axis is aligned with the orbital angular momentum. The angles $\theta_L$ and $\phi_L$ are the polar angles of $\bm{L}$, while $\theta_A$ and $\phi_A$ are the polar angles of $\bm{S}_A$. 
\end{itemize}

%%%%%%%%%%%%%%%%%%%%%%%%%%%%%%%%%%%%%%%%%%%%%%%%%%%%%%%%%%
\section{Spin and Angular Momentum Evolution}
\label{L-S}

A quasicircular binary system consisting of generic spinning compact objects is subject to spin-orbit and spin-spin interactions that force all angular momenta to precess. Averaging over one orbit\footnote{Orbit-averaging should be well justified provided there is a clean separation between the orbital and the precessional time scales, as is the case in the early inspiral.}, the precession equations governing the conservative evolution of the orbital and spin angular momenta are~\cite{Thorne:1984mz,springerlink:10.1007/BF00756587,Buonanno:2005xu}\footnote{The precession equations used here are only strictly valid for BHs, as for NSs they acquire additional terms describing the quadrupole moment of the bodies~\cite{PhysRevD.57.5287}. However, the extra terms are degenerate with the spins~\cite{Krishnendu:2017shb} and difficult to measure.}
\begin{align}
\duvec{L}&= \left\{ \left(2+\frac{3}{2}q\right) - \frac{3}{2}\frac{ v}{\eta}\left[\left( \bm{S}_2 + q\bm{S}_1\right)\cdot \bm{\hat{L}}\right] \right\}  v^6 \left(\bm{S}_1\times \bm{\hat{L}}\right) \nn
\\
&+  \left\{\! \left(2+\frac{3}{2q}\right) - \frac{3}{2}\frac{ v}{\eta}\left[\left( \bm{S}_1 + \frac{1}{q}\bm{S}_2\right)\!\cdot\! \bm{\hat{L}}\right]\!\right\}   v^6 \left( \bm{S}_2\times \bm{\hat{L}} \right)
\nn \\
&+ {\cal{O}}( v^{7}) , \label{Ldot}
\\
\dvec{S}_1 &= \left\{ \eta \left(2+\frac{3}{2}q\right)  - \frac{3 v}{2} \left[\left(q \bm{S}_1 + \bm{S}_2\right) \cdot \bm{\hat{L}}\right]
\right\}
 v^5  \left(\bm{\hat{L}}\times \bm{S}_1 \right)
\nn
\\
&+ \frac{ v^6}{2} \bm{S}_2\times \bm{S}_1 + {\cal{O}}( v^{7})
\label{S1dot},
\\
\dvec{S}_2 &\!= \!\left\{\! \eta \left(2+\frac{3}{2q}\right)  - \frac{3 v}{2} \left[ \left(\frac{1}{q}\bm{S}_2 + \bm{S}_1\right) \!\cdot\! \bm{\hat{L}}\right] \!\right\} 
 v^5  \left(\bm{\hat{L}}\times \bm{S}_2\right) \nn
\\
&+ \frac{ v^6}{2 }\bm{S}_1\times\bm{S}_2 + {\cal{O}}(v^{7})
\label{S2dot}.
\end{align}

Radiation reaction drives the evolution of the magnitude of the orbital angular momentum, leaving the magnitude of the spin angular momenta unaltered to our current knowledge of the PN expansion and ignoring all energy and angular momentum flux through BH horizons~\cite{Poisson:2004cw,Chatziioannou:2012gq}. The magnitude $L$ is related to the evolution of the orbital frequency $\omega$, and the PN expansion parameter $ v$, leading to
\begin{align}
\dot{ v}&=\frac{ v^9}{3}\frac{1}{\sum_{n=0}^{7}\left[ g_n+3 g^{\ell}_n \ln({ v})\right] v^n}.\label{vdot}
\end{align}
The coefficients $\{ g_{n}, g^{\ell}_{n}\}$ are functions of the symmetric mass ratio and inner products of the angular momenta, given in Appendix~\ref{vdot-coeff}.

Equations~\eqref{Ldot}-\eqref{S2dot} describe the {\emph{conservative dynamics}}, while Eq.~\eqref{vdot} describes the {\emph{dissipative dynamics}}. The former models the spin-spin and spin-orbit interactions, that change only the direction of $\bm{L}, \bm{S_1}$ and $\bm{S_2}$. We use only the leading PN order expressions in each interaction\footnote{Spin-orbit corrections can be found in~\cite{Faye:2006gx,Bohe:2012mr}, spin-spin in~\cite{Bohe:2015ana}, and spin-cubed in~\cite{Marsat:2014xea}.}. We do not use higher PN order corrections because the spin-spin and spin-cubed terms have not been fully calculated for generic precessing orbits yet~\cite{Bohe:2015ana,PhysRevD.80.044010}. In principle, we could have included the spin-orbit corrections. However, as explained later, our solution makes use of a certain quantity~\cite{Racine:2008qv} that is conserved by the leading-order in spin-orbit and spin-spin interactions precession equations. If we use partial precession equations (including spin-orbit but not spin-spin corrections) it is not clear if we can modify this quantity so that it remains conserved. Once the spin-spin and spin-cubed terms have been fully calculated we can revisit this problem.

The dissipative dynamics govern the GW frequency evolution by changing the 
magnitude of the Newtonian orbital angular momentum $L=\eta/ v$. This 
equation is known to $2.5$PN order in all spin 
interactions~\cite{PhysRevD.80.044010}, $4$PN in linear-in-spin 
terms~\cite{PhysRevD.80.044010,PhysRevD.80.024002,PhysRevD.84.064041,Bohe:2012mr,Bohe:2013cla,Marsat:2013caa} and $22$PN order in the point particle limit, 
neglecting spins and BH absorption 
effects~\cite{Tanaka:1997dj,Shibata:1994jx,Mino:1997bx,Fujita:2012cm}. In our 
analysis we keep terms in Eq.~\eqref{vdot} to $3.5$PN order 
since this is the highest complete PN order, ignoring spin-spin terms. In this case, we
can easily include partial PN terms in radiation reaction to 
make the evolution more accurate. When the 3PN 
spin-spin term has been fully calculated for precessing 
orbits~\cite{Bohe:2015ana} we can include it in our model. 

Conservative and dissipative equations evolve on distinct time scales. The former evolve on the \emph{precession time scale} 
\begin{align}
T_{\pr} & \equiv \frac{|\bm{S}_1|}{|\dot{\bm{S}}_1|} \sim  v^{-5} ,
\end{align}
while the later evolve on the \emph{radiation reaction time scale}
\begin{align}
T_{\rr} & \equiv \frac{ v}{\dot{ v}} \sim  v^{-8}.
\end{align}
The ratio $T_{\pr}/T_{\rr} \sim  v^3$ is a small quantity in the inspiral and thus a natural expansion parameter.

Recently, Kesden et~al.~\cite{Kesden:2014sla} found an exact solution to the 
precession equations [Eqs.~\eqref{Ldot}-\eqref{S2dot}] ignoring radiation 
reaction [Eq.~\eqref{vdot}]. This solution can be used to ``precession-average'' 
the full precession equations with radiation reaction (analogously to 
orbit-averaging). The final precession-averaged equations depend only on 
quantities that vary on the radiation reaction time scale, and can be 
numerically integrated with a larger step size~\cite{Gerosa:2015tea}. 

Here we take a different approach. Rather that precession-averaging Eqs.~\eqref{Ldot}-\eqref{S2dot} and numerically accounting for Eq.~\eqref{vdot}, we make explicit use of the fact that $T_{\pr}/T_{\rr} \sim  v^3$ to solve the precession equations analytically. We use a perturbation theory technique known as multiple scale analysis (MSA) and treat radiation reaction as a slowly-evolving perturbation on top of precession. This approach allows us to find a solution to the full set of Eqs.~\eqref{Ldot}-\eqref{vdot} as an expansion in $T_{\pr}/T_{\rr}$.

%%%%%%%%%%%%%%%%%%%%%%%%%%%%%%%%%%%%%%%%%%%%%%%%%%%%%%%%%%
\section{Analytic solution to the precession equations without radiation reaction}
\label{precession}

Ignoring radiation reaction, the precession equations can be solved  analytically by 
making use of certain conserved quantities of the system. Below we review and complete the solution first presented in~\cite{Kesden:2014sla}. 

A precessing binary has a total of 9 degrees of freedom arising from the $3$ components of $3$ Newtonian vectors $(\bm{L}, \bm{S_1}, \bm{S_2})$. The precession equations lead to $7$ conserved quantities, reducing the degrees of freedom to $2$. Of the remaining degrees of freedom, one is associated with the choice of a coordinate system, while the other corresponds to a dynamical quantity that changes with 
time. This dynamical quantity is chosen to be the magnitude of the total spin angular momentum 
$S=|\bf{S_{1}} + \bf{S_{2}}|$. 

The conserved quantities are $\bm{\lambda}\equiv(S_1,S_2,L,J,\bm{\hat{J}},\xi)$: 
the magnitudes of the spin angular momenta, the magnitude of the orbital 
angular momentum, the magnitude and direction of the total angular momentum, and 
the mass weighted effective spin~\cite{Racine:2008qv}
\be
\xi \equiv (1+q)\bm{S}_1 \cdot \bm{\hat{L}}  +(1+q^{-1})\bm{S}_2 \cdot \bm{\hat{L}} \label{xi-def}.
\ee
In the effective-one-body formalism, $\xi$ corresponds to the projection of the spin angular momentum of the body at the center of mass onto the orbital angular momentum. Once the system is allowed to evolve under radiation reaction, $S_1, S_2$, and $\xi$ are still conserved, while $L,J$ and $\bm{\hat{J}}$ evolve on the radiation reaction time scale.

In the remainder of this section we use these $7$ conserved quantities to geometrically solve for the $9$ components of the angular momenta as a function of $S$ in a specific coordinate system. We then complete the solution for the angular momenta as a function of time by solving a differential equation to determine $S(t)$.

%------------------------------------------------------------------------------
\subsection{Precession in a noninertial frame}
\label{noninertial}

\begin{figure}[htbp]
\begin{center}
\includegraphics[width=\columnwidth,clip=true]{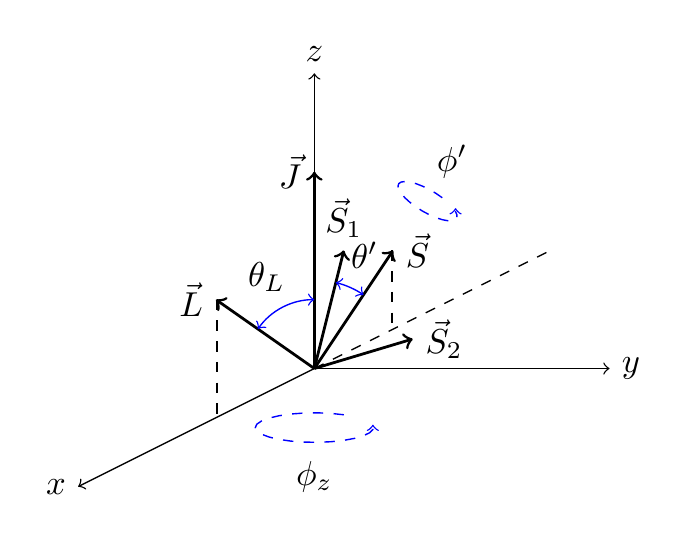}
\caption{\label{config}  Initial configuration of the angular momenta in a noninertial frame precessing around $\bm{\hat{z}}$.}
\end{center}
\end{figure}

The identification of $\bm{\hat{J}}$ as a conserved quantity suggests a 
coordinate frame where $\bm{\hat{z}}=\bm{\hat{J}}$ (see Fig.~\ref{config}). We 
further pick the $x$ and $y$ axes to be precessing around $\bm{\hat{z}}$ (a 
noninertial frame), following the precession of the orbital angular momentum 
which is chosen to be in the $x-z$ plane, at an angle
\be
\cos{\theta_L} = {\bm{\hat{J}}} \cdot {\bm{\hat{L}}} = \frac{J^2 + L^2 -S^2}{2 J L}, \label{thetaL-def}
\ee
from the $\bf{\hat{z}}$ axis.
This allows us to express $\bm{L}$ as
\be
\bm{L}(S;\bm{\lambda}) = L \; [\sin{\theta_L}, 0, \cos{\theta_L}] \label{L-sol}.
\ee
The total spin angular momentum then is
\be
\bm{S}(S;\bm{\lambda}) = \bm{J} - \bm{L} = [-L \sin{\theta_L}, 0, J - L\cos{\theta_L}].
\ee

In another frame with $\bm{\hat{z}'}=\bm{\hat{S}}, \bm{\hat{y}'}=\bm{\hat{y}}$, and $\bm{\hat{x}'}=\bm{\hat{y}'}\times\bm{\hat{z}'}$, we define angles $(\theta',\phi')$ (see Fig.~\ref{config}) such that
\be
\bm{S_1}' = S_1 [\sin{\theta'} \cos{\phi'}, \sin{\theta'} \sin{\phi'}, \cos{\theta'} ].\label{S1primed-sol}
\ee
Using the definition of $\xi$ given in Eq.~\eqref{xi-def} we get
\begin{align}
\cos{\theta'} &=  {\bm{\hat{S}_{1}}} \cdot {\bm{\hat{S}}} = \frac{S^2+S_1^2-S_2^2}{2 S S_1}\label{costhetaprime-def},
\\
\cos{\phi'} &\!=\!\left\{ \!(J^2-L^2-S^2)\!\left[S^2(1+q)^2 - (S_1^2 - S_2^2)(1-q^2)\right] \right.\nn
\\
&\left.- 4 q S^2 L \xi\right\}/[(1-q^2) A_1 A_2 A_3 A_4],\label{cosphiprime-def}
\end{align}
where
\begin{align}
A_1 &= \sqrt{ J^2 - (L - S)^2}, \\
A_2 &= \sqrt{ (L + S)^2 - J^2}, \\
A_3 &= \sqrt{ S^2 - (S_1 - S_2)^2}, \\
A_4 &= \sqrt{ (S_1 + S_2)^2 - S^2}.
\end{align}
In the original unprimed system
\be
\bm{S_1}(S;\bm{\lambda})  ={\mathbb{R}}(\uvec{y},\theta_S)  \bm{S_1}' , \label{S1-sol}
\ee
where $\mathbb{R}(\uvec{y},\theta_S)$ is a rotation around $\uvec{y}$ by an angle $\theta_{S}$ and 
\be
\cos{\theta_S} = \bm{\hat{S}} \cdot \bm{\hat{J}} =  \frac{J^2 + S^2 -L^2}{2 J S}.\label{costhetaS-def}
\ee
Once we have $\bm{S_1}$ in the original unprimed system, then 
\be
\bm{S_2}(S;\bm{\lambda}) = \bm{J} - \bm{L} - \bm{S_1} \label{S2-sol}. 
\ee
Equations~\eqref{L-sol},~\eqref{S1-sol}, and~\eqref{S2-sol} determine the angular momenta in a noninertial frame as a function of $S$ up to the sign of $\sin\phi'$ in Eq.~\eqref{S1-sol}, which we will tackle shortly.

At this point, the various orbital angular momenta have been written in a noninertial frame in terms of $S$ using purely geometrical arguments. The evolution equation of $S$ can be derived from Eqs.~\eqref{Ldot}-\eqref{S2dot}:
\be
\left( \frac{d S^2}{dt} \right)^2 = - A^2 \left(S^6 + B S^4 + C S^2 
+D \right),\label{dSdt-prel}
\ee
where the coefficients $A, B, C, D$ depend only on quantities that change on the 
radiation reaction time scale. Their explicit form is given in 
Appendix~\ref{coeff}. The roots of the polynomial on the right-hand 
side of Eq.~\eqref{dSdt-prel} have a simple interpretation. When $S^2$ is equal to one of the roots, its 
derivative is zero. Therefore, two of the roots are the maximum $S_+^2$ and the 
minimum $S_-^2$ of $S^2$. The third root $S_3^2$ does not 
correspond to any physically interesting scenario; in fact, it is negative for 
most systems{\footnote{In the most generic case, a third order polynomial with 
real coefficients can have complex roots. However, we argue that this is an 
unphysical scenario. Unless two of the roots are real, $S^2$ will increase or 
decrease with no bound. If two roots of a third order polynomial with real 
coefficients are real, then the third root must be real too.}}. 

Making explicit use of the roots of the polynomial, we can rewrite Eq.~\eqref{dSdt-prel} as
\be
\left(\frac{d S^2}{dt}\right)^2 = -A^2 (S^2 - S_+^2) (S^2 - S_-^2) (S^2 - S_3^2).\label{dSdt-final}
\ee
The solution to this equation is
\be
S^2 = S_+^2 + (S_-^2 - S_+^2) \, \rm{sn}^2({\psi,m}) \label{Ssq-sol}
\ee
where $\rm{sn}$ is a Jacobi elliptic function (see Sec. 16 
of~\cite{Abramowitz:1970as} for a detailed introduction to the Jacobi elliptic 
functions, and~\cite{2007arXiv0711.4064B} for a physics-oriented approach), 
$\psi$ is its phase, and $m \in [0,1]$. When $m=0$, $\text{sn}$ reduces to a 
sine, while for $m=1$ it gives a hyperbolic tangent. The period of $S^2$ 
 is $2 K(m)$, where $K(m)$ is the complete elliptic integral of the 
first kind. The phase and the parameter $m$ are given by
\be
\frac{d \psi}{d t} = \frac{A}{2} \sqrt{S_+^2 - S_3^2} \label{psidot}
\ee
and
\be
m=\frac{S_+^2-S_-^2}{S_+^2-S_3^2}\,. \label{m-def}
\ee
Clearly, this solution requires that $S_{+}^{2} \neq S_{3}^{2}$, which is almost always the case because $S_{+}^{2}$ and $S_{3}^{2}$ are defined to be the largest and smallest roots respectively. The only possible case when $S^{2}_{3} = S^{2}_{+}$ is when $S^{2}_{+} = S^{2}_{-}$, but then $S^{2}$ is constant in the first place and there is no precession. The phase $\psi$ can be obtained by noticing that $\dot{\psi}$ is constant if we ignore radiation reaction, so that
\be
\psi = \frac{A}{2} \sqrt{S_+^2 - S_3^2}\, t.\label{psisol}
\ee

The final ingredient we need in order to have a complete expression for all 
angular momenta as function of time in a non-inertial frame precessing around 
$\bm{\hat{z}}$ is the sign of $\sin{\phi'}$. Equation~\eqref{S1primed-sol} 
implies that
\be
\text{sign}(\sin{\phi'}) = \text{sign}(\bm{S_1} \cdot \bm{y}'),
\ee
which after some algebra can be shown to be equivalent to 
\begin{align}
\text{sign}(\sin{\phi'}) &= \text{sign}\left[(\bm{\hat{L}} \times \bm{S_1}) \cdot \bm{S_2}\right] = \text{sign}\left(-\frac{dS^2}{dt}\right)\nn
\\
&= \text{sign}\left[\rm{sn}({\psi,m}) \rm{cn}({\psi,m})\right],
\end{align}
where $ \rm{cn}({\psi,m})$ is another Jacobi elliptic function and in the last equality we have used Eq.~\eqref{Ssq-sol}.

%------------------------------------------------------------------------------
\subsection{Precession in an inertial frame}
\label{inertial}

All angular momenta so far have been expressed in a noninertial frame that precesses 
around $\bm{\hat{J}}$. An Euler rotation of 
$\bm{L}, \bm{S_1}$, and $\bm{S_2}$ around $\bm{\hat{z}}$ by some angle $\phi_z$  and 
substitution into the precession equations yields~\cite{Kesden:2014sla}
\begin{align}
\frac{d \phi_z} {d t} &\equiv \Omega_z = \frac{J}{2} v^6 \left\{ 1+ \frac{3}{2\eta}\left(1-\xi  v\right) \right. \nn
\\
&\left. - \frac{3(1+q)}{2qA_1^2A_2^2}\left( 1- \xi  v\right)  \left[ 4(1-q)L^2 (S_1^2 - S_2^2) \right. \right. \nn
\\
&\left. \left. - (1+q)(J^2-L^2-S^2)(J^2-L^2-S^2-4\eta  L \xi)\right]\right\}.\label{phizdot-def}
\end{align}

The precession angle $\phi_z$ changes on the precession time scale through $S$ and on the radiation-reaction time scale through $J$ and $L$. We recast it in the form
\begin{align}
\frac{\dot{\phi}_z}{J} &= a + \frac{c_0 + c_2 \,\text{sn}^2(\psi, m) + c_4 \,\text{sn}^4(\psi, m)}{d_0 + d_2\, \text{sn}^2( \psi, m ) + d_4 \,\text{sn}^4( \psi, m)},\label{phizdot}
\end{align}
where $a$, the $d_i$'s and the $c_i$'s are quantities that evolve on the radiation reaction time scale only. Their explicit form is given in Appendix~\ref{coeff}. Now $\dot{\phi_z}$ can be integrated exactly in the absence of radiation reaction to give
\begin{align}
 \frac{\phi_z}{J} &= A_{\phi}\frac{\psi}{\dot{\psi}} + iB_{\phi} \frac{F[i \sinh^{-1}{\left(\text{sc}(\psi,m)\right)},1-m] }{\dot{\psi}}\nn
 \\
 & +i C_{\phi} \frac{\Pi[n_c, i \sinh^{-1}{\left(\text{sc}(\psi,m)\right)},1-m]}{\dot{\psi}} \nn
 \\
 &+ i D_{\phi}\frac{ \Pi[n_d, i \sinh^{-1}{\left(\text{sc}(\psi,m)\right)},1-m]}{\dot{\psi}}   ,\label{phiz-norr}
\end{align}
where $\dot{\psi}$ is given by Eq.~\eqref{psidot}, $F$ is the elliptic integral of the first kind, $\Pi$ is the elliptic integral of the third kind, and $\text{sc}$ is a Jacobi elliptic function. The quantities $A_{\phi}, B_{\phi}, C_{\phi}, D_{\phi}, n_c, n_d$ are functions of $\{a,c_i,d_i\}$, and they are constant in the absence of radiation reaction. They are given in Appendix~\ref{coeff}. 

This concludes the solution to the precession equations in the absence of radiation reaction in a frame where $\uvec{J}=\uvec{z}$. In summary, at some initial time:
\begin{itemize}
\item The orbital angular momentum  $\bm{L}$ is given by Eq.~\eqref{L-sol}, which depends on the angle $\theta_{L}$ given in Eq.~\eqref{thetaL-def}. The latter depends on $S$, which varies on the precession timescale as described in Eq.~\eqref{Ssq-sol}; 
\item The spin angular momentum of the heavier body $\bm{S_{1}}$ is given in Eq.~\eqref{S1-sol}, which depends on the angle $\theta_{S}$ given in Eq.~\eqref{costhetaS-def} as well as on $\mb{S_{1}'}$ given in Eq.~\eqref{S1primed-sol} in terms of the angles $(\theta',\phi')$ of Eqs.~\eqref{costhetaprime-def} and~\eqref{cosphiprime-def}. All of these depend on $S$, which again is described by Eq.~\eqref{Ssq-sol};
\item The spin angular momentum of the lighter body $\mb{S_{2}}$ is given by Eq.~\eqref{S2-sol}, which depends on $\mb{L}$ and $\mb{S_{1}}$ described above.
\end{itemize}
The full precessional motion of these angular momenta in an inertial frame is obtained by rotating them around $\uvec{z}$ by $\phi_z$, given in Eq.~\eqref{phiz-norr}.

%%%%%%%%%%%%%%%%%%%%%%%%%%%%
\section{Addition of radiation reaction}
\label{radiationreaction}

The exact solution to the precession equations obtained in the previous section 
is valid only in the absence of radiation reaction. 
The problem of including radiation reaction admits a 
perturbative solution owing to its two distinct time scales: radiation reaction 
unfolds on a much longer time scale than precession. This natural separation of 
timescales allows us to treat radiation reaction as a slow perturbation of the 
more rapid precession, a technique formally known as multiple scale 
analysis~\cite{Bender}. 

In MSA, every quantity is expanded in the ratio of the two distinct timescales. 
In our case, we expand in the ratio of the precessional time scale $T_{\pr}$ to the 
radiation reaction time scale $T_{\rr}$; radiation reaction is a 1.5PN effect on 
top of precession. This is not the first application of MSA to the precession 
problem. In fact, the precession equations we started with 
are orbit-averaged, which would be the first term in an MSA expansion about the 
ratio of the fast orbital time scale to the precession time scale. 

%------------------------------------------------------------
\subsection{Choice of an inertial frame}
\label{frame-rr}

\begin{figure*}[htbp]
\begin{center}
\vspace{-0.6cm}
\includegraphics[width=\columnwidth,clip=true]{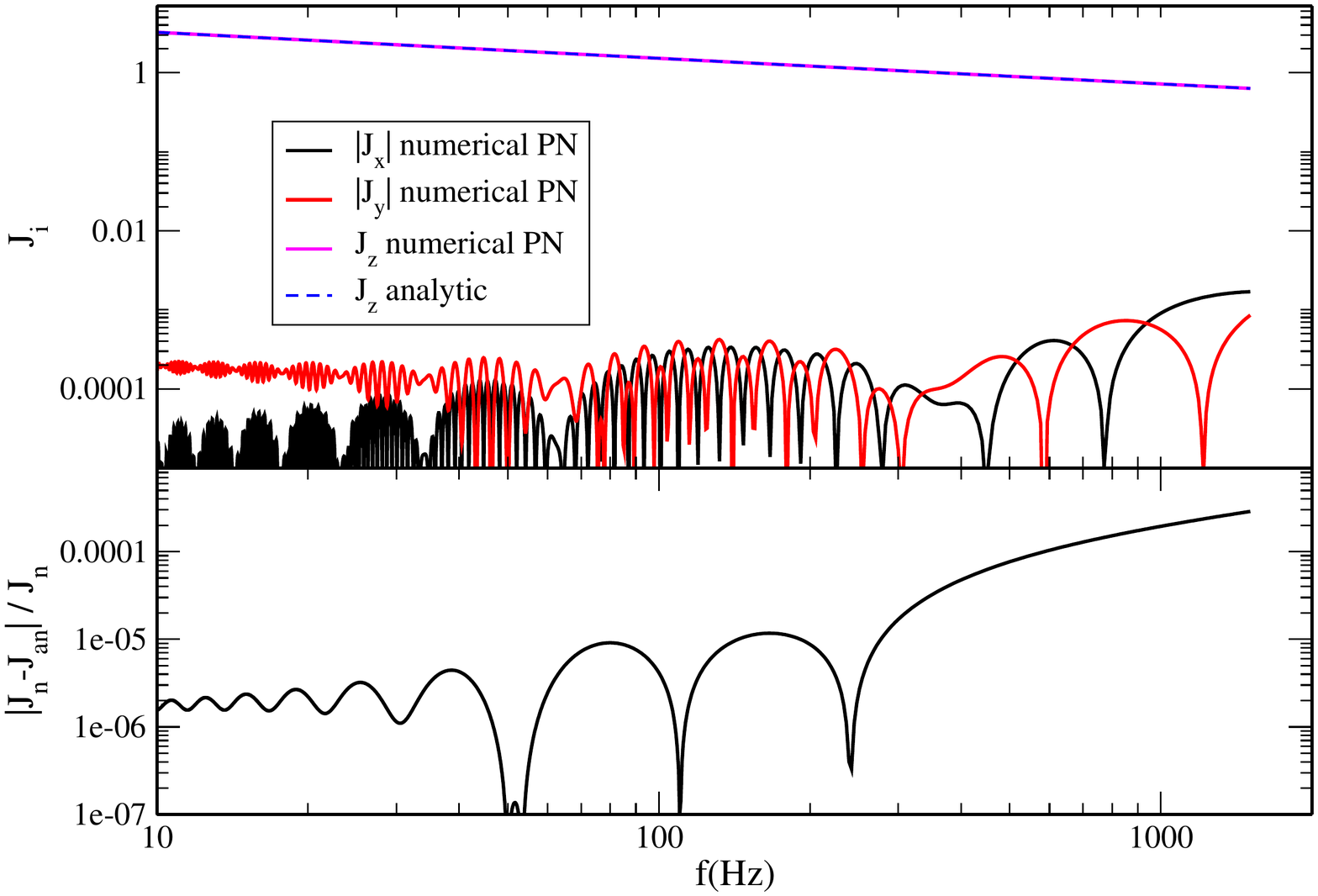}
\includegraphics[width=\columnwidth,clip=true]{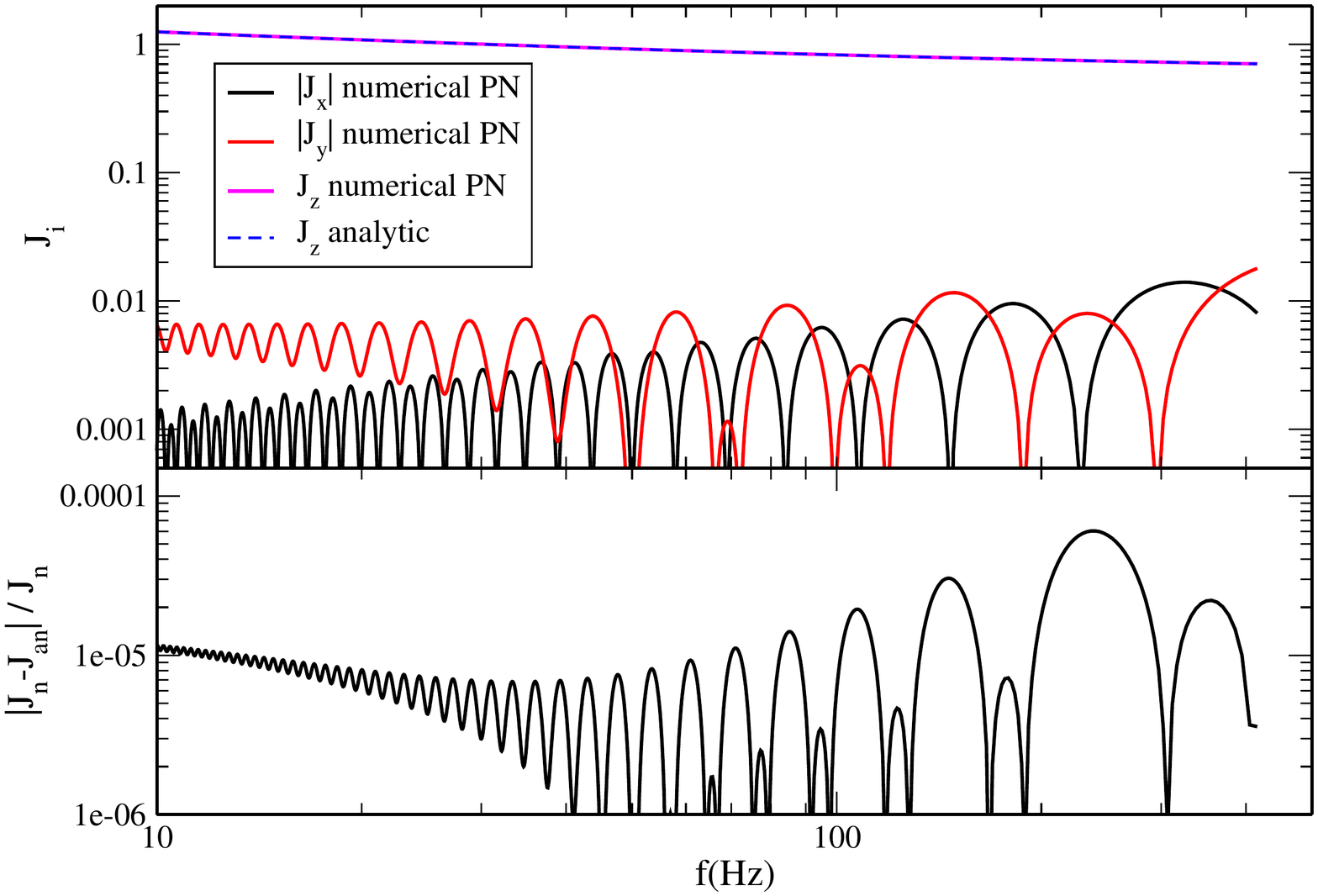}\\
\vspace{-0.6cm}
\includegraphics[width=\columnwidth,clip=true]{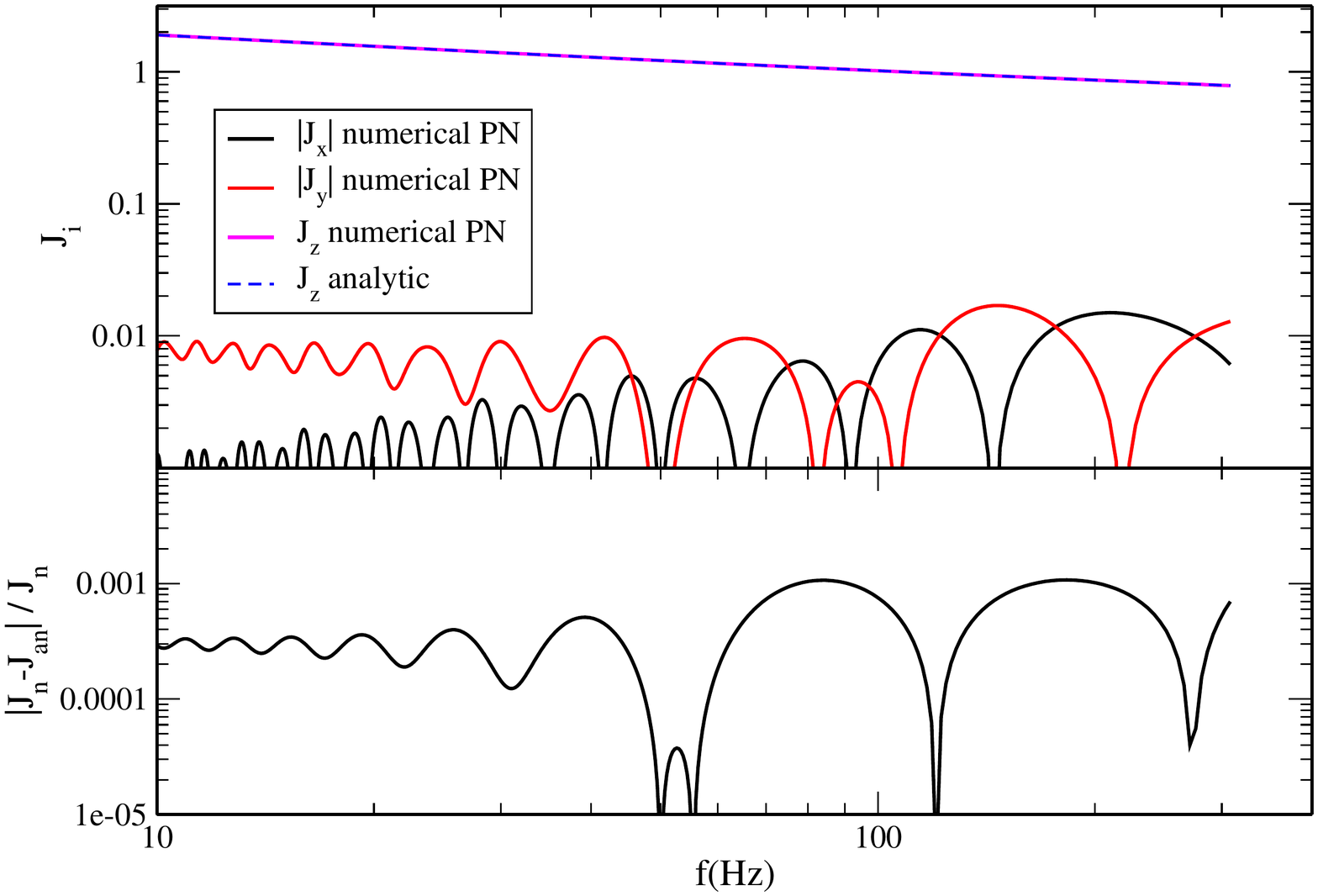}
\includegraphics[width=\columnwidth,clip=true]{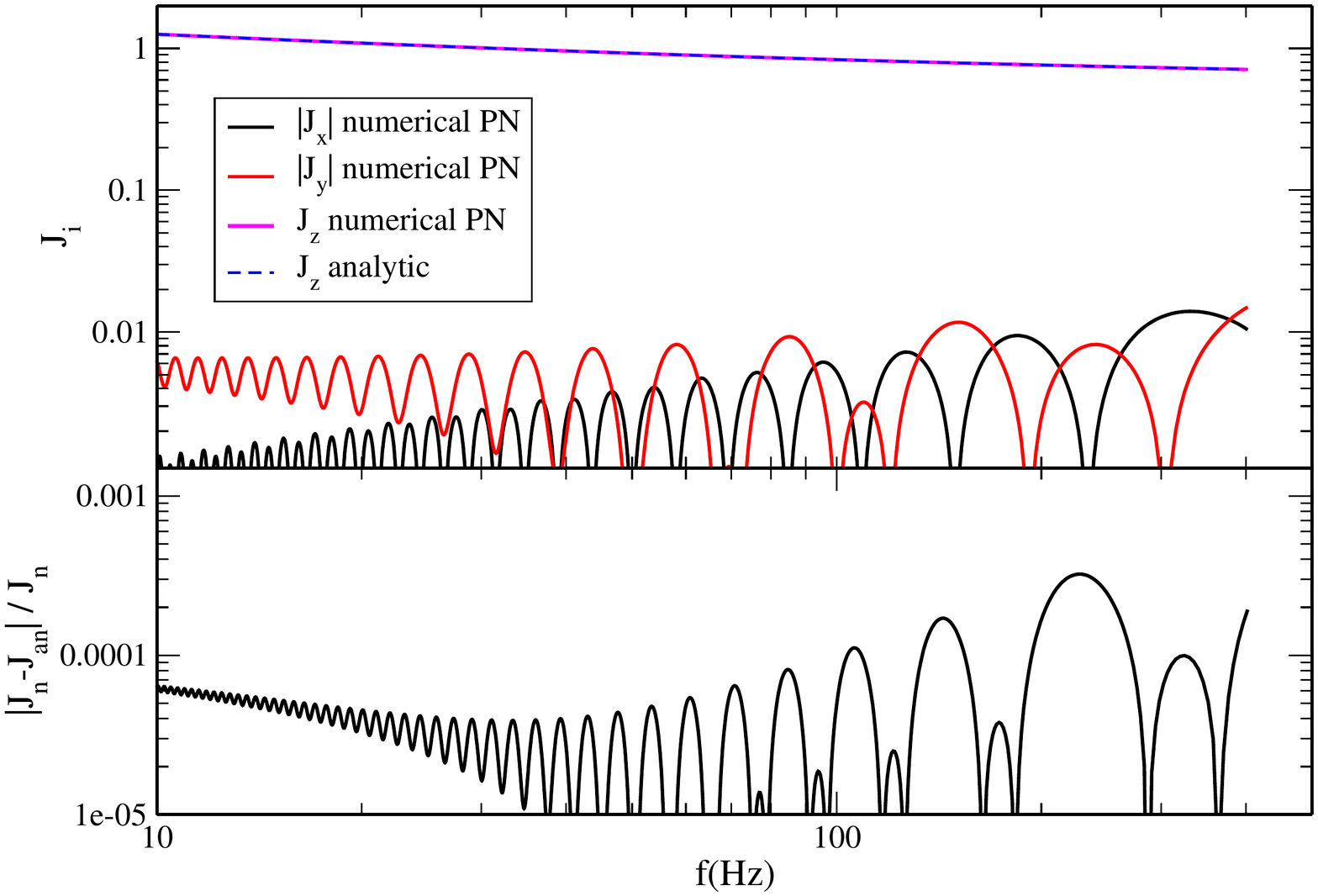}
\vspace{-0.5cm}
\caption{\label{fig:Jcomparisons} (Top Panel) Comparison between the numerical PN and 
the analytic components of the total angular momentum as a function of the GW frequency for the NSNS (Top Left), the BHNS (Top Right), the BHBH (Bottom Left), and the HSNSBH (Bottom Right) system of Table~\ref{table:systems}. (Bottom Panel) Fractional error between the magnitude of 
the total angular momentum obtained numerically and analytically.}
\end{center}
\end{figure*}

The precession solution of Sec.~\ref{precession} was built around the assumption that $\uvec{J}$ is conserved and aligned with $\uvec{z}$. Our first task when adding radiation reaction is to check whether this remains true. If it does, then the functional form of Eqs.~\eqref{L-sol},~\eqref{S1-sol}, and~\eqref{S2-sol} holds, since they were derived solely on geometrical arguments.

Radiation reaction does not strictly conserve the direction of the total angular momentum. However, it has been argued~\cite{Apostolatos:1994mx} that in the context of simple precession $(\bm{S_2}=0)$ the variation of $\uvec{J}$ in a precession cycle averages out. Here we show that this is approximately true for generic precession as well~\cite{Buonanno:2002fy}.

Equations~\eqref{Ldot}-\eqref{S2dot} imply 
\begin{align}
 \dvec{J} &= \dot{L} \uvec{L},
\end{align}
and after some algebra we can rewrite this as
\begin{align}
\dot{\uvec{J}} & = \frac{\dot{L}}{J L} \bm{L} - \frac{\dot{J}}{J^2}\bm{J}.
\end{align}
Averaging over $\phi_z$ we find
\begin{align}
\precphi{\dot{J}_x} & = \precphi{\frac{\dot{L}}{J} \sin{\theta_L} \cos{\phi_z}},
\\
\precphi{\dot{J}_y} & = \precphi{\frac{\dot{L}}{J} \sin{\theta_L} \sin{\phi_z}},
\\
\precphi{\dot{J}_z} & = 0.
\end{align}
This averaging induces an error in $\dot{\uvec{J}}$ that is ${\cal{O}}(T_{\pr}/T_{\rr})$, or $\dot{\uvec{J}}-\langle\dot{\uvec{J}}\rangle\sub{$\phi_z$} \sim  v^3$. At this order, we can treat $\dot{L}$ as a constant, since the spin couplings in Eq.~\eqref{vdot} first enter at ${\cal{O}}( v^3)$. They are therefore of the same order as the averaging error, and can be neglected.

Working to this order we have
\begin{align}
\precphi{\dot{J}_{x,y}} & \sim \precphi{\sin{\theta_L} \cos{\phi_z}}\nn
\\
& \sim \precphi{ \sqrt{1-\left(\frac{J^2+L^2-S^2}{2JL}\right)^2} \cos{\phi_z}}  .
\end{align}

Since $L \sim {\cal{O}}( v^{-1})$, $J \sim {\cal{O}}( v^{-1})$, and $S \sim {\cal{O}}( v^{0})$, a PN expansion yields schematically
\begin{align}
\precphi{\!\dot{J}_{x,y}\!} &\!\! \sim   \precphi{\cos{\phi_z}} \!+ v^{2} \!\precphi{S^2 \cos{\phi_z}} \!+ {\cal{O}}( v^{4}).
\end{align}
The first term vanishes, while the second is of higher PN order  and we neglect it. This situation is different from simple precession. In the latter the averaging out of $\dot{\uvec{J}}$ is exact, while here it requires a PN expansion. We therefore expect this result to become less and less accurate as the binary approaches merger.

The above calculation implies that $\langle\dot{\uvec{J}}\rangle\sub{$\phi_z$}=0$; radiation reaction changes the magnitude of $\bm{J}$ while leaving its direction approximately constant. The components $J_x$ and $J_y$ are expected to oscillate with an amplitude much smaller than $J_z$ without exhibiting any secular growth. Figure~\ref{fig:Jcomparisons} tests the validity of this statement. We select 4 systems with typical parameters as expected for NSNS, BHNS, BHNS, and HSNSBH binaries (see Table~\ref{table:systems}) and plot the components of $\bm{J}$ obtained by numerically solving Eqs.~\eqref{Ldot}-\eqref{vdot} as a function of the GW frequency $f$. In all cases $J_x$ and $J_y$ are at least $2$ orders of magnitude smaller that $J_z$ and oscillate around $0$, with no signs of secular growth.

Based on this result we can build a solution to the precession equations including radiation reaction in the inertial frame introduced in Sec.~\ref{precession}. That is, we neglect any variation in the direction of $\uvec{J}$ and align it with $\uvec{z}$. This choice of frame automatically means that the functional form of Eqs.~\eqref{L-sol},~\eqref{S1-sol}, and~\eqref{S2-sol} for the orbital and spin angular momenta respectively is still valid, since they were derived on purely geometric arguments. On the contrary, any quantity that was derived based on Eqs.~\eqref{Ldot}-\eqref{S2dot} needs to be revisited and recalculated by taking Eq.~\eqref{vdot} into account. This involves the remaining $5$ conserved quantities of precession $(S_1, S_2, L, J, \xi)$, Eq.~\eqref{dSdt-final} for the magnitude of the total spin angular momentum, and Eq.~\eqref{phizdot} for the precession angle.

 %----------------------------------------------------------------------
\subsection{Constants of the precessional motion}
\label{Jmag}

In principle, the constants of the precessional motion need not remain constant when radiation reaction is invoked.
The magnitudes of the two spin angular momenta $S_1$ and $S_2$, and the mass weighted effective spin $\xi$ remain constant under radiation reaction to the PN order we work here and ignoring horizon absorption. The magnitude of the orbital angular momentum $L$ is updated by definition through $L= \eta/ v$. The magnitude of the total angular momentum $J$ depends on $L$ and also changes under radiation reaction. The evolution equation for $J$ averaged over one period of $S(t)$ is~\cite{Kesden:2014sla}
\begin{align}
 \precav{\frac{dJ}{dL}} &= \frac{J^2 + L^2 - \precav{S^2} }{2JL} .
\end{align}
 This can be integrated exactly to yield
\begin{align}
J^2 &= L^2 + \frac{2c_1}{\eta} L - L \int \frac{\precav{S^2}}{L^2} dL,\label{J-almost}
\end{align}
where here and in what follows $J$ is approximated by its precession 
average, and $c_1$ is an integration constant.
As we will show below, $\precav{S^2}$ is constant when ignoring high-order PN effects, 
and the integral of Eq.~\eqref{J-almost} can be calculated to give
\begin{align}
 J^2 &= L^2 + \frac{2c_1}{ v} + \precav{S^2} + {\cal{O}}( v).\label{Jsol}
\end{align}

The quantity $\precav{S^2}$ can be computed from Eq.~\eqref{Ssq-sol}:
\begin{align}
 S^2_{\av}\! \equiv\! \precav{S^2}\! =\! \frac{1}{m} \bigg[\! ( m - 1) 
S^2_+ + S^2_-+ \frac{E(m)}{K(m)} \left( S^2_+ \!-\! S^2_- \right)\! \bigg],
\end{align}
where $K(m)$ and $E(m)$ are the complete elliptic integrals of the 
first and second kind respectively. 
PN expanding $S^2_+$ and $S^2_-$ around their initial value we find
\begin{align}
 S^2_\pm &= S^2_{\pm, 0} + \ord{ v}, \qquad
 S^2_3 = \ord{ v^{-2}}, 
\end{align}
which together with Eq.~\eqref{m-def} yields
\begin{align}
m = \ord{ v^2},
\end{align}
and
\begin{align}
S^2\sub{av} &= \frac{1}{2} \left( S^2_{+,0} + S^2_{-,0} \right) + \ord{ v^2}.\label{Save-def}
\end{align}
In the above expressions $S^2_{\pm, 0}$ are the roots computed from the initial 
conditions.
 
Combining the result for $J$ obtained here and Sec.~\ref{frame-rr} where we 
justified keeping $\uvec{J}$ aligned with $\uvec{z}$, our analytic approximation 
for the total angular momentum is 
\be
\bm{J} = [0,0,J]\label{J-sol}.
\ee
To verify that this approximate $\bm{J}$ stays close to the numerical PN solution we plot it in Fig.~\ref{fig:Jcomparisons} as a function of the GW frequency for our three study systems. The analytic $J_x$ and $J_y$ are identically zero, so we omit them. The bottom panel shows the fractional error in the magnitude of the total angular momentum when approximated by Eq.~\eqref{Jsol}. The maximum discrepancy in the magnitude $J$ is of $\ord{10^{-2}}$ in those particular examples indicating both that Eq.~\eqref{Jsol} is accurate and that setting $J_x$ and $J_y$ equal to zero is justified.

 %----------------------------------------------------------------------------------------
\subsection{Magnitude of the total spin angular momentum}
\label{Smag}

\begin{figure*}[htbp]
\begin{center}
\vspace{-0.6cm}
\includegraphics[width=\columnwidth,clip=true]{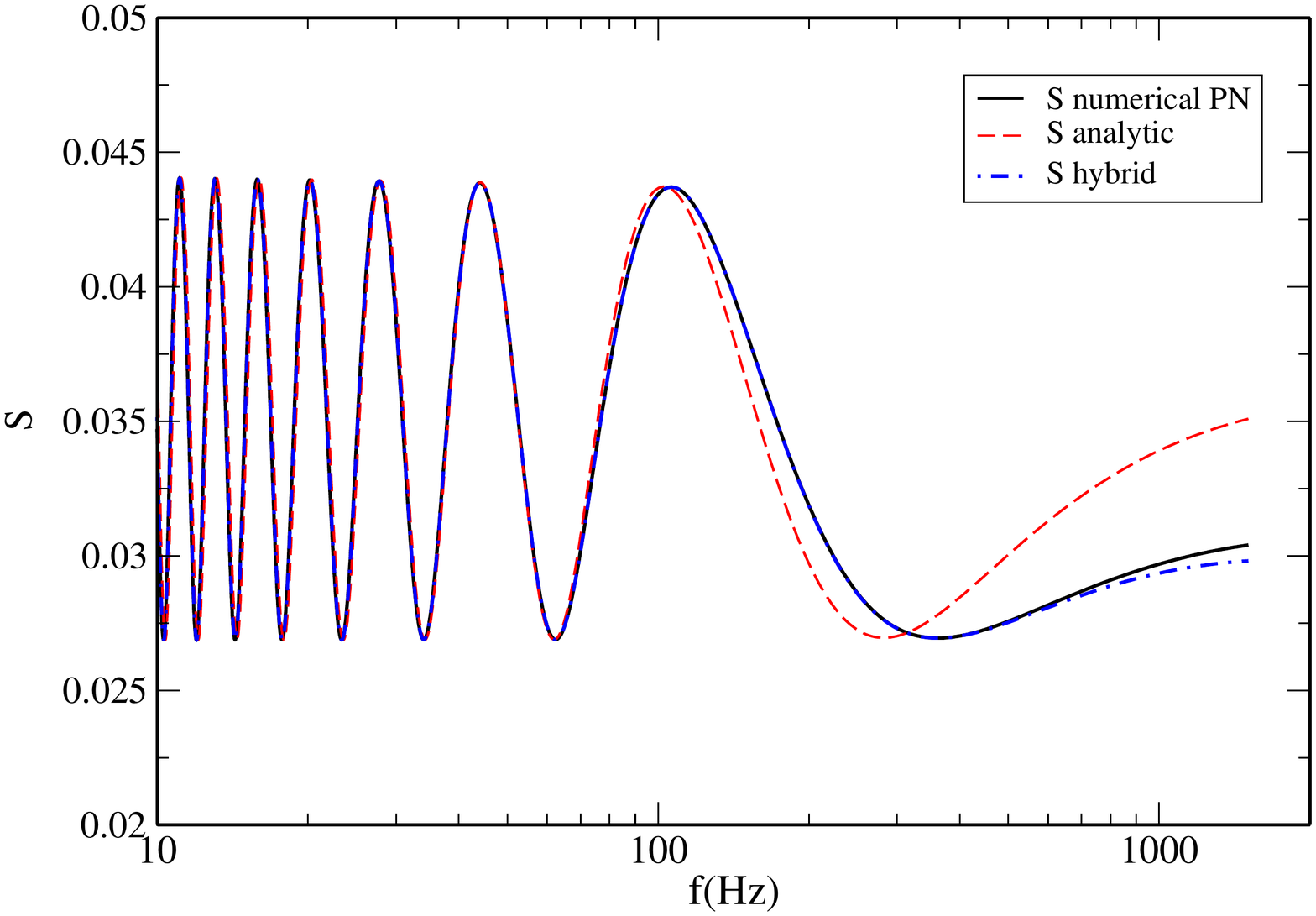}
\includegraphics[width=\columnwidth,clip=true]{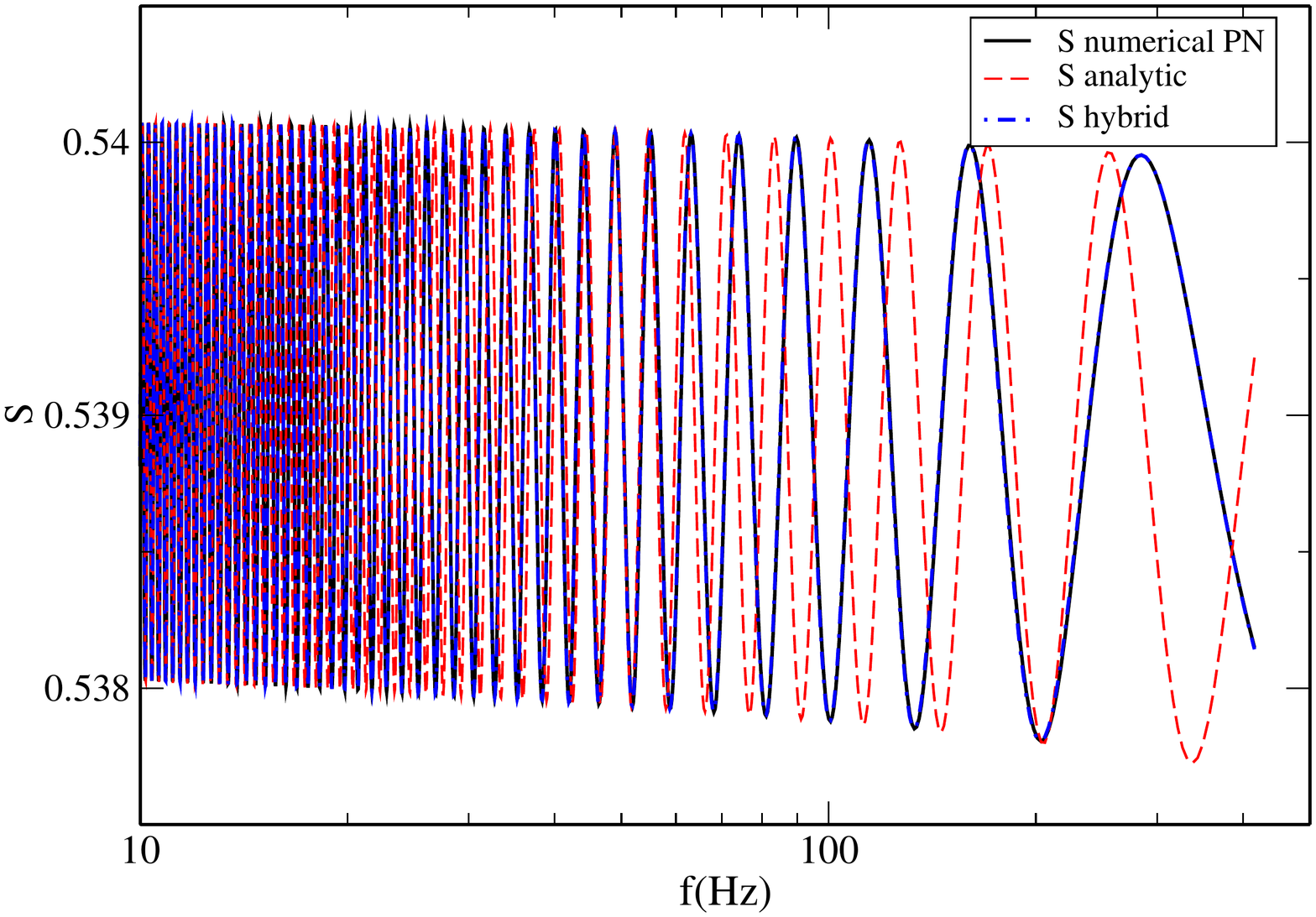}\\
\vspace{-0.5cm}
\includegraphics[width=\columnwidth,clip=true]{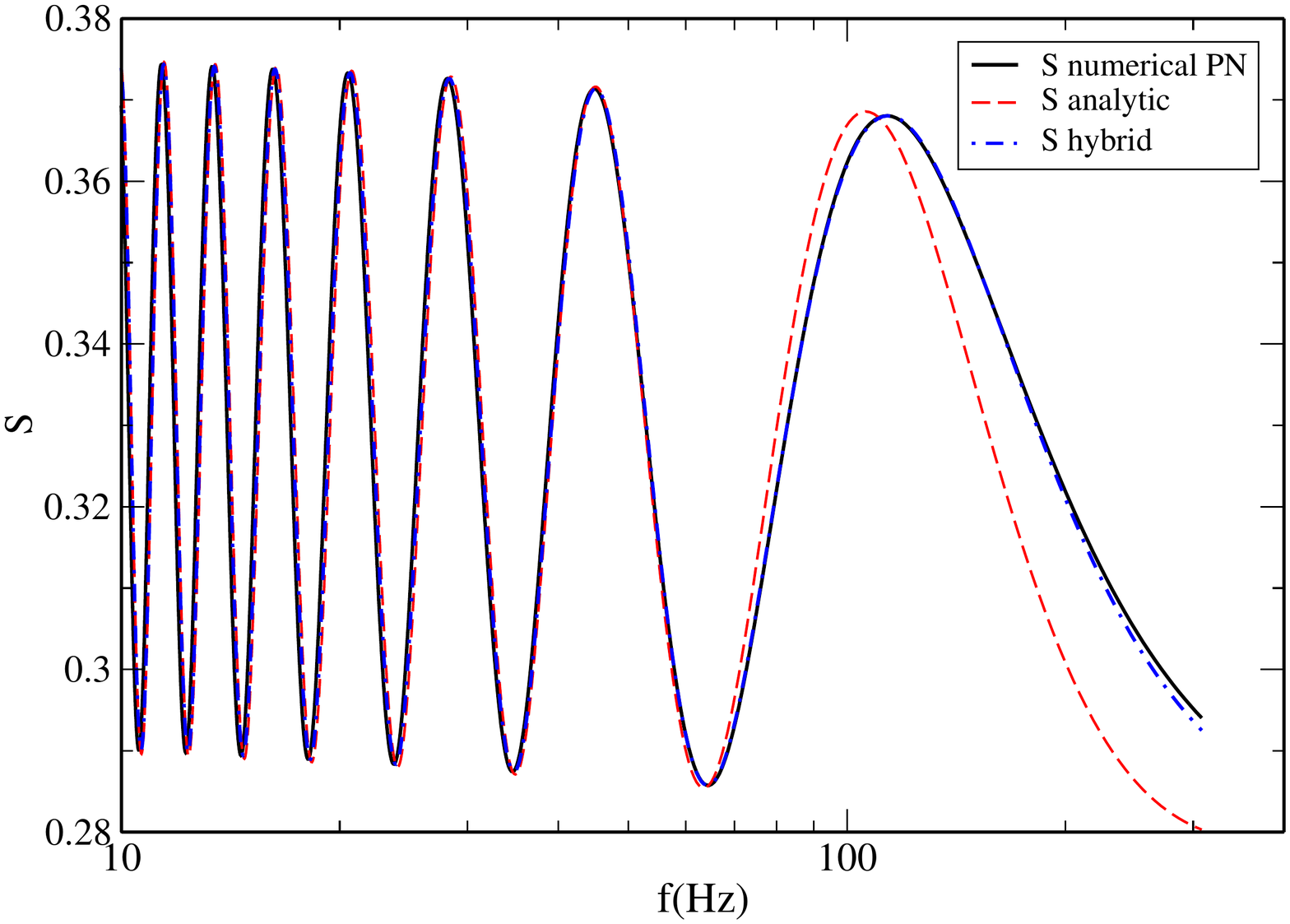}
\includegraphics[width=\columnwidth,clip=true]{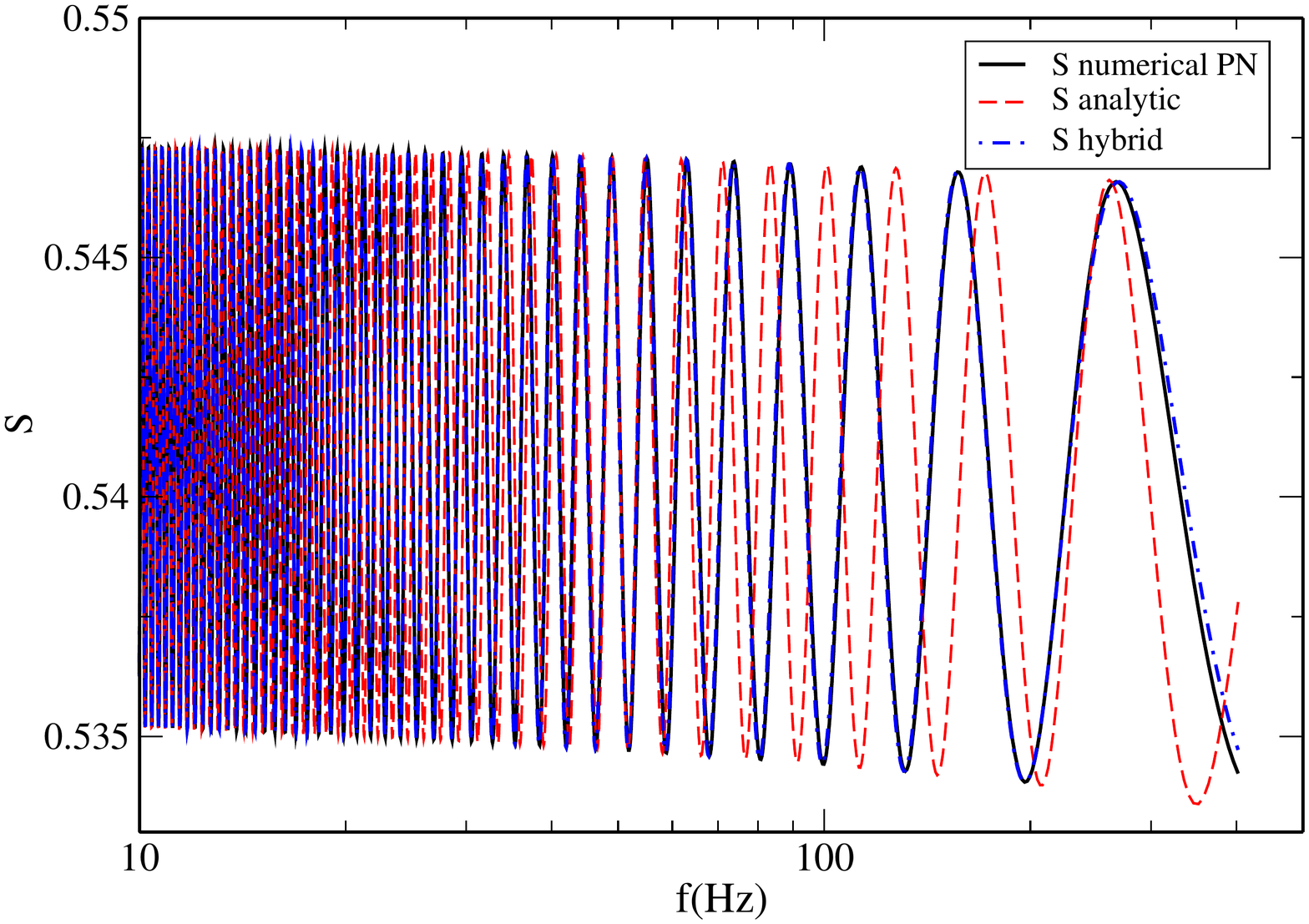}
\vspace{-0.5cm}
\caption{\label{fig:Scomparisons} Comparison between the numerical PN (black solid), 
the analytic (red dashed), and the hybrid (blue dot-dashed) magnitude of the total spin angular momentum as a function of the 
GW frequency for the NSNS (Top Left), the BHNS (Top Right), the BHBH (Bottom Left), and the HSNSBH (Bottom Right) system of Table~\ref{table:systems}. }
\end{center}
\end{figure*}

Once radiation reaction is included, Eq.~\eqref{dSdt-final} for the magnitude of the total spin angular momentum needs to be solved with MSA. 
We first explicitly separate the time scales by writing $S^2(t) = S^2(t_{\pr}, t_{\rr})$, where $t_{\pr}$ 
denotes variation on the precession time scale, while $t_{\rr} = \epsilon t_{\pr}$ 
denotes variations on the radiation reaction time scale, with $\epsilon$ 
a bookkeeping parameter. 

Expanding $S^2$ as 
\begin{align}
 S^2(t_{\pr}, t_{\rr}) = \sum_{n \geq 0}  \epsilon^n S^2_n(t_{\pr}, t_{\rr}). 
\end{align}
and substituting this expression into Eq.~\eqref{dSdt-final}, at leading order in $\epsilon$, 
we recover Eq.~\eqref{dSdt-final} for 
$S^2_0(t_{\pr}, t_{\rr})$ with the time derivative taken on the precession time scale $t_{\pr}$:
\begin{align}
\left(\frac{\p S^2_0}{\p t_{\pr}}\right)^2 &= -A^2(t_{\rr}) [S^2_0(t_{\pr}, t_{\rr}) - S_+^2(t_{\rr})] 
\\
&\times [S^2_0(t_{\pr}, t_{\rr}) - S_-^2(t_{\rr})]\nn [S^2_0(t_{\pr}, t_{\rr}) - S_3^2(t_{\rr})].
\end{align}
The solution to this differential equation is similar to Eq.~\eqref{Ssq-sol}, except that quantities that were previously constant are now promoted to functions of $t_{\rr}$:
\begin{align}
 S^2_0 &= S^2_+(t_{\rr}) + \left[ S^2_-(t_{\rr}) - S^2_+ (t_{\rr}) \right] 
\text{sn}[\psi(t_{\pr}, t_{\rr}), m(t_{\rr})],\label{S-sol}
\end{align}
where $S^2_+(t_{\rr})$, $S^2_-(t_{\rr})$, and $m(t_{\rr})$ now depend on time through $L(t_{\rr})$ and $J(t_{\rr})$. 

The angle $\psi(t_{\pr}, t_{\rr})$ satisfies 
\begin{align}
\frac{d \psi}{d t} = \frac{A(t_{\rr})}{2} \sqrt{S_+^2(t_{\rr}) - S_3^2(t_{\rr})}.\label{psidot-rr}
\end{align}
where we keep terms of $\ord{\epsilon}$ by taking the derivative with respect to $t$ rather than $t_{\pr}$. We can integrate this equation using a PN 
integration, i.e. expanding it in powers of $ v$ and integrating term by 
term. The result is
\begin{align}
\psi &= \psi_0 -\frac{3  g_0}{4 }\delta m \, v^{-3} \left(1+ \psi_1 v + 
\psi_2  v^2 \right), \label{psi_rr}
\end{align}
where $\psi_0$ is an integration constant, and the 
constants $\psi_1, \psi_2$ are given in Appendix~\ref{coeff-psi}. 
We find that expanding Eq.~\eqref{psi_rr} to relative 1PN order 
suffices.

We test this solution for $S$ in Fig.~\ref{fig:Scomparisons} by 
plotting the numerical PN, analytic, and hybrid magnitude of the total spin angular momentum $S$ 
as a function of the GW frequency for the $4$ systems we study. The hybrid $S$ is obtained through Eq.~\eqref{S-sol} but with a numerical solution to Eq.~\eqref{psidot-rr}. For all systems, the amplitude of $S$ shows excellent agreement with the numerical PN
results, which is controlled by the roots $S^2_+$ and $S^2_-$. For the NSNS and 
BHBH systems, the analytic phase $\psi$ also shows very good agreement with the 
numerical PN result, although the dephasing for the BHNS and HSBHNS systems is about 2 cycles. 
However, both systems are dominated by the spin of the BH, making the motion 
close to that of simple precession; the variation in $S$ is very small 
as demonstrated by the scale of the $y$ axis of the right panels of 
Fig.~\ref{fig:Scomparisons} and this dephasing should not affect the emitted waveform considerably. 

On the other hand, the phase of the hybrid $S$ is always in excellent agreement 
with the numerical solution, indicating that if we do indeed need an improved 
solution in the future\footnotemark \ we can obtain it by carrying out the 
expansion of Eq.~\eqref{psi_rr} to higher order.

\subsection{Precession angle}
\label{phizsol}

The final quantity that needs to be recalculated to account for radiation 
reaction is the precession angle. Its derivative, given in 
Eq.~\eqref{phizdot}, depends both on the precession and the radiation reaction 
time scale, so it requires a MSA treatment. 

We write
\begin{align}
\frac{d\phi_z}{dt} &= \Omega_z[S(t),L(t),J(t)] = \Omega_z[S(t_{\pr},t_{\rr}), L(t_{\rr}), J(t_{\rr})],\label{phiz-full}
\end{align}
and expand the precession angle as
\be
\phi_z(t_{\pr},t_{\rr}) = \epsilon^{-1} \phi_{z,-1}(t_{\pr},t_{\rr}) + \phi_{z,0}(t_{\pr},t_{\rr}) + 
\ord{\epsilon}.
\ee
The reason $\phi_z$ includes a term of $\ord{\epsilon^{-1}}$ is because the 
binary precesses even in the absence of radiation reaction.

Solving Eq.~\eqref{phiz-full} order by order in $\epsilon$, we find to 
$\ord{\epsilon^{-1}}$
\be
\frac{1}{\epsilon}\frac{\p \phi_{z,-1}}{\p t_{\pr}} = 0\,,
\ee
which means $\phi_{z,-1}=\phi_{z,-1}(t_{\rr})$. 
To next order, we find 
\be
\frac{\p \phi_{z,-1}}{\p t_{\rr}} + \frac{\p \phi_{z,0}}{\p t_{\pr}} = 
\Omega_z(t_{\pr}, t_{\rr})\,, \label{order-e}
\ee
and averaging over $t_{\pr}$  we find
\be
\frac{d \phi_{z,-1}}{d t_{\rr}} = \precav{\Omega_z} (t_{\rr}),\label{phizm1-def}
\ee
where we set $\precav{ \p \phi_{z,0}/\p t_{\pr} }= 0$ to cancel 
secular terms.
Equation~\eqref{phizm1-def} can be solved with a PN 
integration. Going back to Eq.~\eqref{order-e} we get
\be
\frac{\p \phi_{z,0}}{\p t_{\pr}} = \Omega_z(t_{\pr}, t_{\rr}) -  \precav{\Omega_z} (t_{\rr}).\label{eq1}
\ee
Integrating the first term on the right-hand side of Eq.~\eqref{eq1} we recover Eq.~\eqref{phiz-norr} for $\phi_z$ in the absence of radiation reaction. Integrating the second term is straightforward.

The full solution for $\phi_z$ is then
\be
\phi_z = \phi_{z,-1}+  \phi_{z,0} +{\cal{O}}(\epsilon)\label{phiz-sol}, 
\ee
where
\begin{align}
 \phi_{z,-1} &= \int \precav{\Omega_z}\!(t_{\rr}) \; dt_{\rr},\\
\phi_{z,0} &= \int  \Omega_z(t_{\pr}, t_{\rr})\; dt_{\pr} - \int \precav{\Omega_z}\!(t_{\rr}) \; dt_{\pr}.\label{phiz0-def}
\end{align}
The meaning of each term in the MSA expansion is clear. The first term 
$\phi_{z,-1}$ is averaged over the fast (relative
to radiation reaction) precession time scale, and then integrated 
over radiation reaction. The next term $\phi_{z,0}$ is a first 
order correction to this precession averaging.

%------------------------------------------------------------------------
\subsubsection{Leading order MSA}
\label{leading-msa}

 \begin{figure*}[htbp]
\begin{center}
\vspace{-0.6cm}
\includegraphics[width=\columnwidth,clip=true]{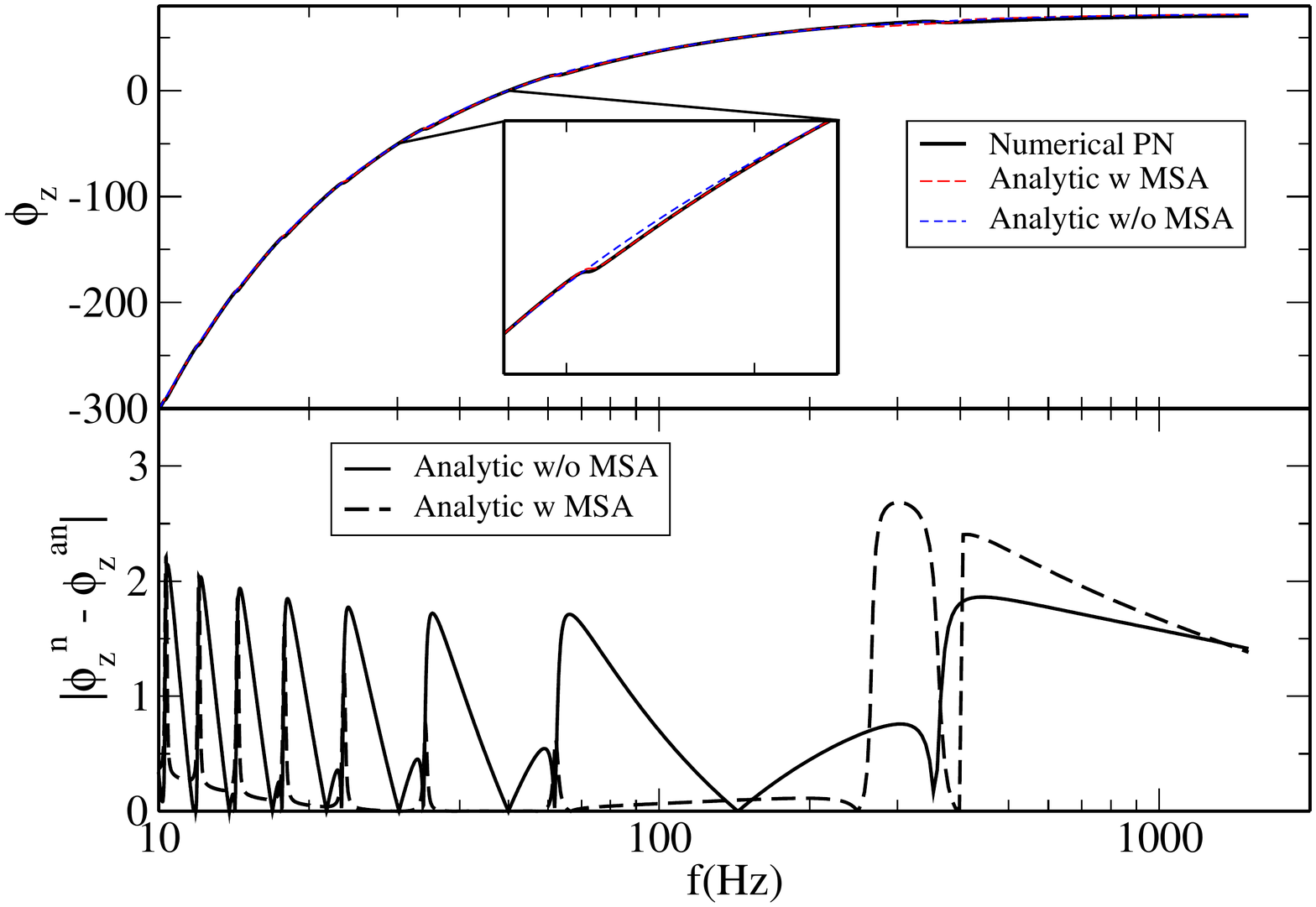}
\includegraphics[width=\columnwidth,clip=true]{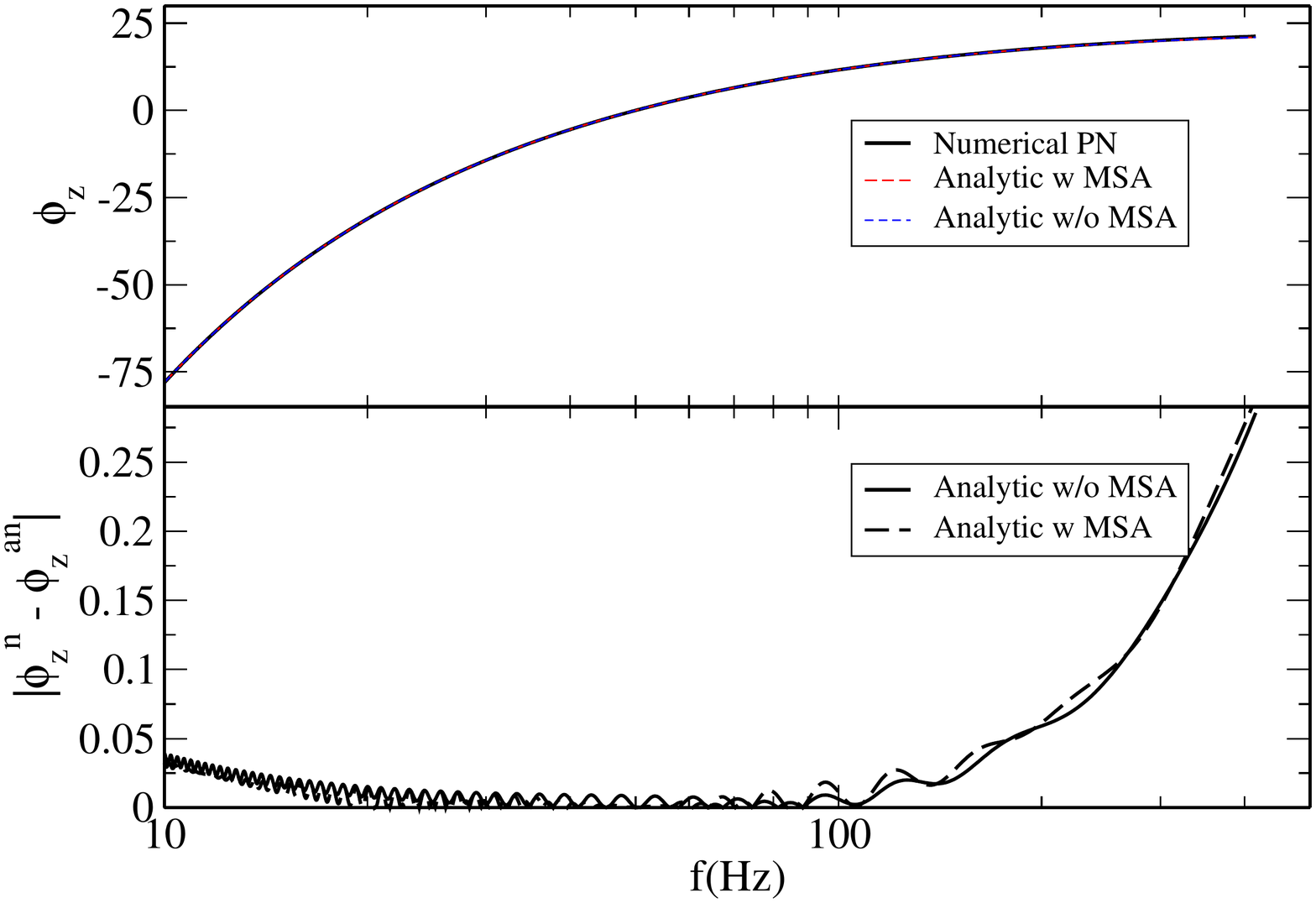}\\
\vspace{-0.5cm}
\includegraphics[width=\columnwidth,clip=true]{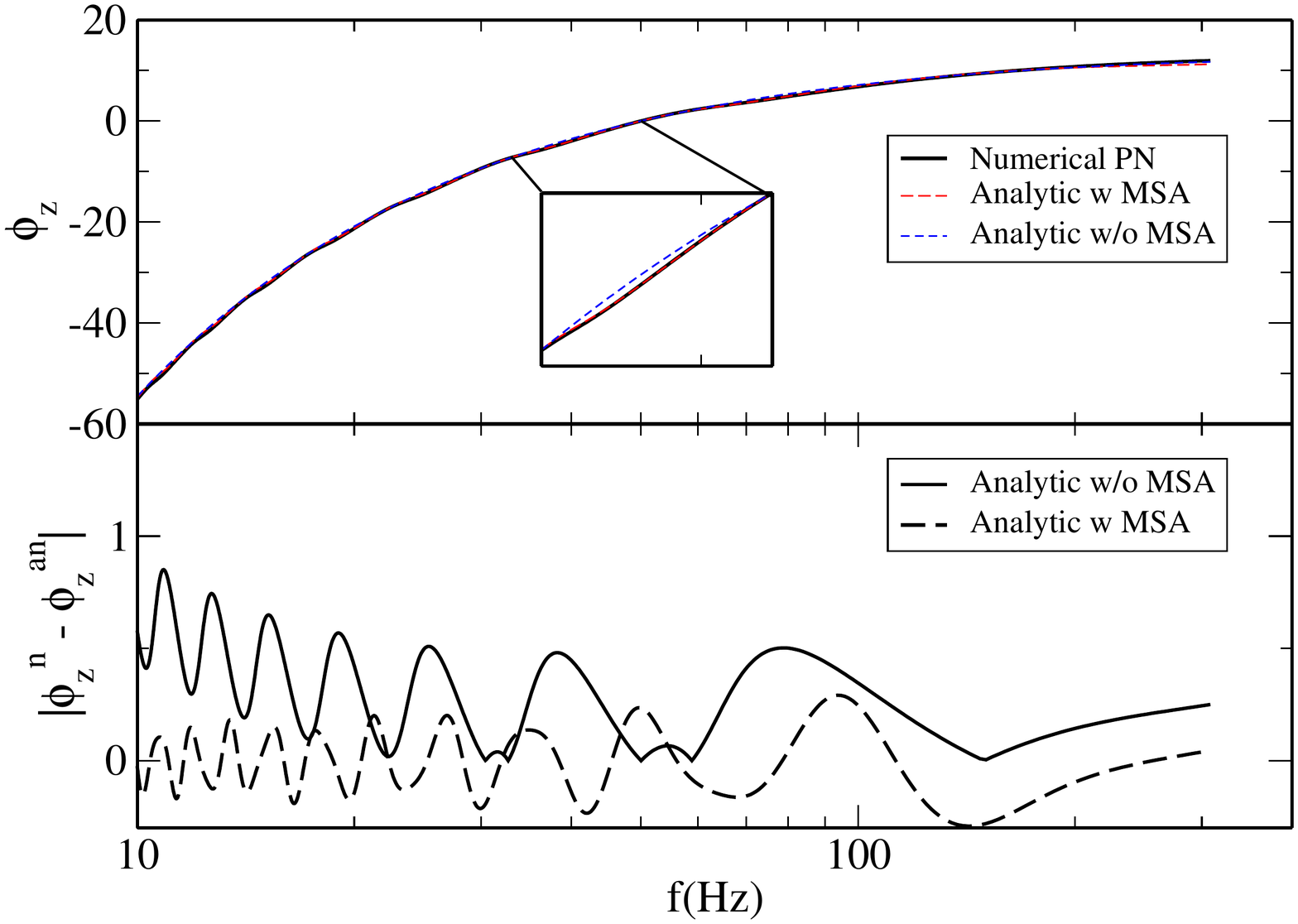}
\includegraphics[width=\columnwidth,clip=true]{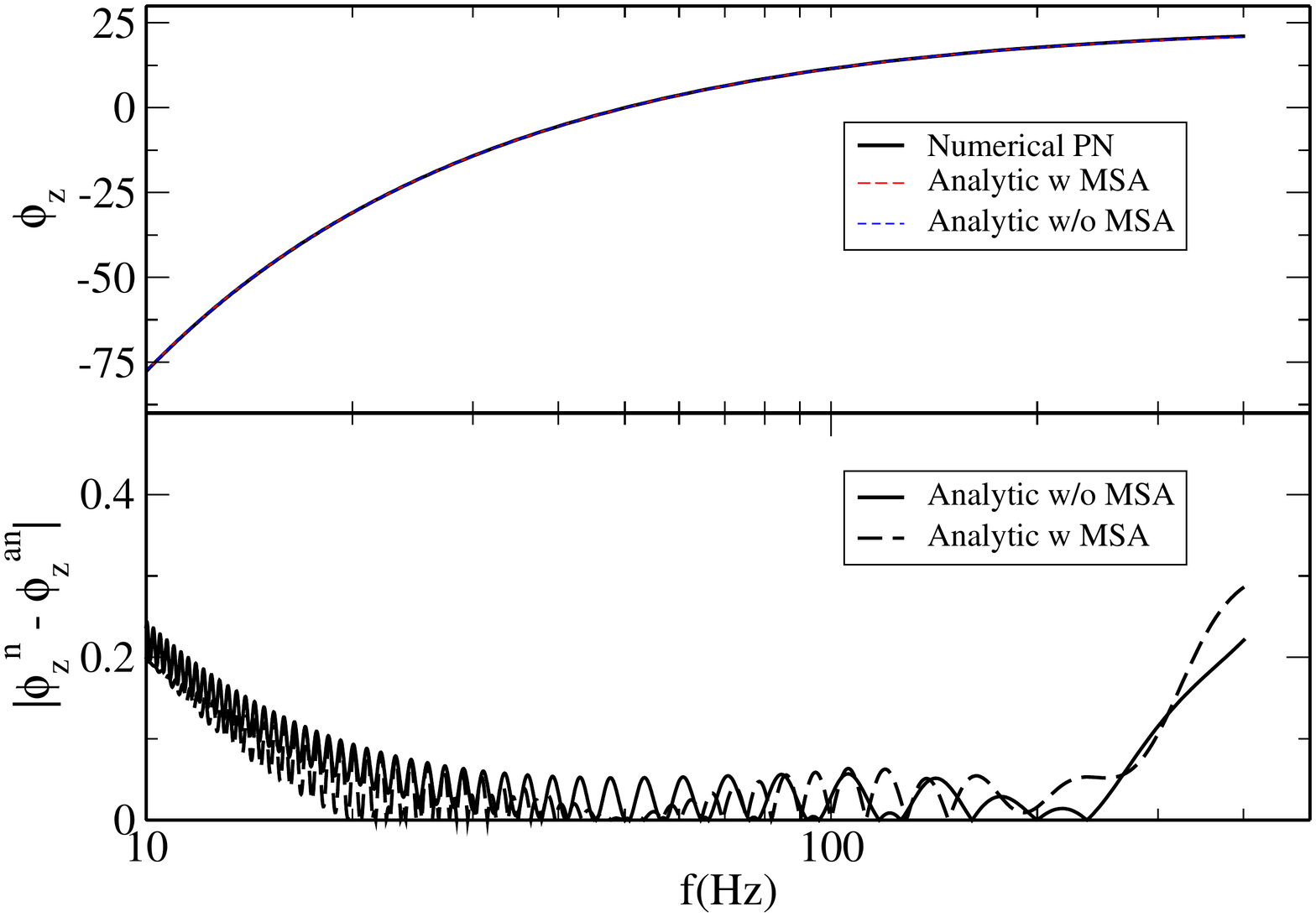}
\vspace{-0.5cm}
\caption{\label{fig:phizcomparisons} (Top Panel) Comparison between the numerical PN and 
the analytic precession phase as a function of the GW frequency for the NSNS (Top Left), the BHNS (Top Right), the BHBH (Bottom Left), and the HSNSBH (Bottom Right) system of Table~\ref{table:systems}. (Bottom Panel) Error in $\phi_z$ with and without the MSA corrections.  }
\end{center}
\end{figure*}

The leading order MSA term is defined in Eq.~\eqref{phizm1-def} which to first order in $\epsilon$ is equivalent to 
\be
\precav{\frac{d \phi_z}{dt}}=\precav{\Omega_z}\label{phizPN}.
\ee
The average of $\Omega_z$ can be obtained by taking the difference between Eq.~\eqref{phiz-norr} 
evaluated at $\psi=0$ and at $\psi=2 K(m)$, where recall that $K(m)$ is the complete elliptic 
integral of the first kind. However, for reasons explained in Appendix~\ref{phizexpressions}, we prefer to use Eq.~\eqref{phizdot} and find an alternative way of calculating $\precav{\dot{\phi}_z }$. We write
\begin{widetext}
\begin{align}
\frac{\dot{\phi}_z}{J}-a &\equiv \dot{\phi}_z^{\text{red.}}= \frac{c_0 + c_2 \,\text{sn}^2(\psi, m) + c_4 \,\text{sn}^4(\psi, m)}{d_0 + d_2 \,\text{sn}^2(\psi, m) + d_4 \,\text{sn}^4(\psi, m)}  \Rightarrow \nn
\\
[d_0 + d_2 \,\text{sn}^2(\psi, m) &+ d_4 \,\text{sn}^4(\psi, m)]\dot{\phi}_z^{\text{red.}}  = c_0 + c_2 \,\text{sn}^2(\psi, m)+ c_4 \,\text{sn}^4(\psi, m) \Rightarrow\nn
\\
 d_0  \precav{\dot{\phi}_z^{\text{red.}} }+ d_2 \precav{\,\text{sn}^2(\psi, 
m)\dot{\phi}_z^{\text{red.}} } &+ d_4 \precav{\,\text{sn}^4(\psi, 
m)\dot{\phi}_z^{\text{red.}} }=  c_0 + c_2\precav{\,\text{sn}^2(\psi, m) } + c_4 
\precav{\,\text{sn}^2(\psi, m)} , \nn
\end{align}
\end{widetext}
\footnotetext{For example, if and when LIGO's sensitivity increases, so will 
its \\ requirement for more accurate waveforms.}
where on the third line we average over precession.
Unfortunately, no closed form expressions exist for $\precav{\,\text{sn}^2(\psi, m) }$ and $\precav{\,\text{sn}^4(\psi, m) }$ for arbitrary m. We can, however, calculate these averages as an expansion 
in $m\ll 1$ since, as already discussed, $m \sim {\cal{O}}( v^2)$. 
We could in principle retain high order in $m$ terms in this expansion, but in practice 
we find that working to leading order in $m$ suffices. Expanding the above expression to leading order in $m \ll 1$, we find
\begin{align}
d_0\precav{\dot{\phi}_z^{\text{red.}} } &+ d_2 \precav{\dot{\phi}_z^{\text{red.}} \sin^2\psi } + d_4 \precav{\dot{\phi}_z^{\text{red.}} \sin^4\psi }\nn 
\\
&=  c_0 + \frac{1}{2}c_2 + \frac{3}{8}  c_4\Rightarrow\nn
\\
d_0  \precav{\dot{\phi}_z^{\text{red.}} }&+ d_2 D_2\precav{\dot{\phi}_z^{\text{red.}} }+ d_4 D_4\precav{\dot{\phi}_z^{\text{red.}} }\nn
\\
&=  c_0 + \frac{1}{2}c_2 +\frac{3}{8} c_4 \Rightarrow\nn
\\
\precav{\dot{\phi}_z}&= J\Big(a+ \frac{c_0 + \frac{1}{2}c_2 + \frac{3}{8} c_4}{d_0 + d_2 D_2+ d_4 D_4}\Big),\label{phizdotfinal}
\end{align}

where we have defined
\begin{align}
D_2&\!\equiv\! \frac{\precav{\dot{\phi}_z^{\text{red.}} \sin^2\psi } }{\precav{\dot{\phi}_z^{\text{red.}} }}\!=\! \frac{\precav{\frac{c_0 + c_2 \sin^2\psi + c_4 \sin^4\psi}{d_0 + d_2\sin^2\psi + d_4 \sin^4\psi}\sin^2\psi } }{\precav{\frac{c_0 + c_2 \sin^2\psi + c_4 \sin^4\psi}{d_0 + d_2\sin^2\psi+ d_4\sin^4\psi}}}\label{D2-def},
\\
D_4&\!\equiv \!\frac{\precav{\dot{\phi}_z^{\text{red.}} \sin^4\psi } }{\precav{\dot{\phi}_z^{\text{red.}} }}\!=\!\frac{\precav{\frac{c_0 + c_2 \sin^2\psi + c_4 \sin^4\psi}{d_0 + d_2\sin^2\psi + d_4 \sin^4\psi}\sin^4\psi } }{\precav{\frac{c_0 + c_2 \sin^2\psi + c_4 \sin^4\psi}{d_0 + d_2\sin^2\psi+ d_4\sin^4\psi}}}\label{D4-def}.
\end{align}
The quantities $D_2$ and $D_4$ are functions of $ v$ and can be calculated exactly. For reasons explained in Appendix ~\ref{phizexpressions} we do not wish to use these full expressions, but rather we keep the quantities $D_2$ and $D_4$ constant and set them equal to their leading PN order expressions.

We can now integrate the right-hand side of Eq.~\eqref{phizdotfinal} by first PN expanding it. However, we find it more convenient to factor 
$J$ out of $ \precav{\Omega_z}$ and PN expand the remaining terms. We do so to avoid 
artificial divergences in the small mass ratio limit arising from expanding 
around essentially $\eta/ v$; see Appendix~\ref{phizexpressions}.
We, then, have to perform an integral of the form 
\be
\phi_{z,-1} = \int \frac{J}{\xi^3} \sum_{n=0}^5 \langle \Omega_z\rangle^{(n)} \, v^n d 
\xi,\label{int-phiz}
\ee
where the coefficients $\langle \Omega_z\rangle^{(n)}$ are given in 
Appendix~\ref{coeff-phiz}. This integral can be directly calculated to give
\begin{align}
\phi_{z,-1}= \sum_{n=0}^5  \langle \Omega_z\rangle^{(n)} \phi_z^{(n)} + 
\phi_{z,-1}^{0},\label{phiz-final}
\end{align}
where $ \phi_z^{(n)}$ are functions given in Appendix~\ref{coeff-phiz} and  
$\phi_{z,-1}^{0}$ is an integration constant.

%------------------------------------------------------------------------
\subsubsection{Correction to MSA}
\label{corr-msa}

The first-order correction to MSA is given in Eq.~\eqref{phiz0-def}. The solution to 
the first integral is Eq.~\eqref{phiz-norr} where we set $m=0$. The second integral is trivial since 
$ \precav{ \Omega_z }$ does not depend on the precession time scale $t_{\pr}$, and the result is $ \precav{ \Omega_z } t_{\pr}$. In that expression, we choose for convenience to substitute $t_{\pr}=\psi/\dot{\psi}$.

Collecting all the elements together, the correction to the precession phase is given by
\begin{align}
\phi_{z,0} &= \frac{C_{\phi} }{\dot{\psi}}\!\frac{\sqrt{n_c}}{n_c-1}\text{arctan}\left[\frac{(1-\sqrt{n_c}) \tan\psi}{1+\sqrt{n_c}  \tan^2\psi} \right]\nn
\\
&+ \frac{D_{\phi}}{\dot{\psi}} \frac{\sqrt{n_d}}{n_d-1}\text{arctan}\left[\frac{(1-\sqrt{n_d}) \tan\psi}{1+\sqrt{n_d}  \tan^2\psi} \right]\label{phiz0},
\end{align}
where $\dot{\psi}$ is given in Eq.~\eqref{psidot}, $\psi$ is given in 
Eq.~\eqref{psi_rr} and $C_{\phi},D_{\phi},n_c$ and $n_d$ are functions of 
$ v$ given in Appendix~\ref{coeff}.

%------------------------------------------------------------------------
\subsubsection{Comparisons}

In Fig.~\ref{fig:phizcomparisons} we plot the numerical PN and analytic solutions for $\phi_z$ with and without the MSA corrections. The small oscillations of the numerical PN phase are reproduced by the analytic phase with MSA corrections. These oscillations are more pronounced for the NSNS and BHBH systems where both spins contribute significantly to the dynamics. The bottom panel shows the error in the precession phase with and without MSA corrections.

%%%%%%%%%%%%%%%%%%%%%%%%%%%%%%%%%%%%%%%%%%%%%%%%%%%%%%%%%%
\section{Building the Waveform}
\label{sua}

Using the solution for the angular momenta described above, we calculate an 
analytic time-domain waveform for 
generic precessing binaries. The gravitational wave signal emitted by 
a precessing binary system as observed in an interferometric detector is~\cite{Blanchet:2014,Apostolatos:1994mx,Arun:2008kb,PhysRevD.84.124011,
Lundgren:2013jla}:
\begin{align}
 h(t) &= F_+h_+ +  F_\times h_\times, 
 \end{align}
 where
 \begin{align}
 F_+ &= \frac{1}{2} \left( 1 + \cos^2 \theta_N' \right) \cos2 \phi_N' \cos 2 
\psi_p \nn\\
&- \cos\theta_N' \sin2\phi_N' \sin 2\psi_p, \\
 F_\times &= \frac{1}{2} \left( 1 + \cos^2 \theta_N' \right) \cos2 \phi_N' \sin 
2 
\psi_p \nn\\
&+ \cos\theta_N' \sin2\phi_N' \cos 2\psi_p ,
\end{align}
are the antenna pattern functions, $h_{+,\times}$ 
are the GW polarization states, $(\theta_N', \phi_N')$ are the polar angles of 
$\uvec{N}$ in a frame tied 
to the arms of the detector with $\uvec{z}'$ the normal to the detector plane, 
and 
$\psi_p$ is given by
\begin{align}
 \psi_p &= \text{arctan} \left[ \frac{ \left( P_N \uvec{J} \right) \cdot 
\uvec{z}'}{\left(\uvec{N} \times \uvec{J} \right) \cdot \uvec{z}'} \right], 
\label{eq:psiL}
\end{align}
where $P_N$ acts as a projection along $\uvec{N}$.

 \begin{figure*}[htbp]
\begin{center}
\vspace{-0.6cm}
\includegraphics[width=\columnwidth,clip=true]{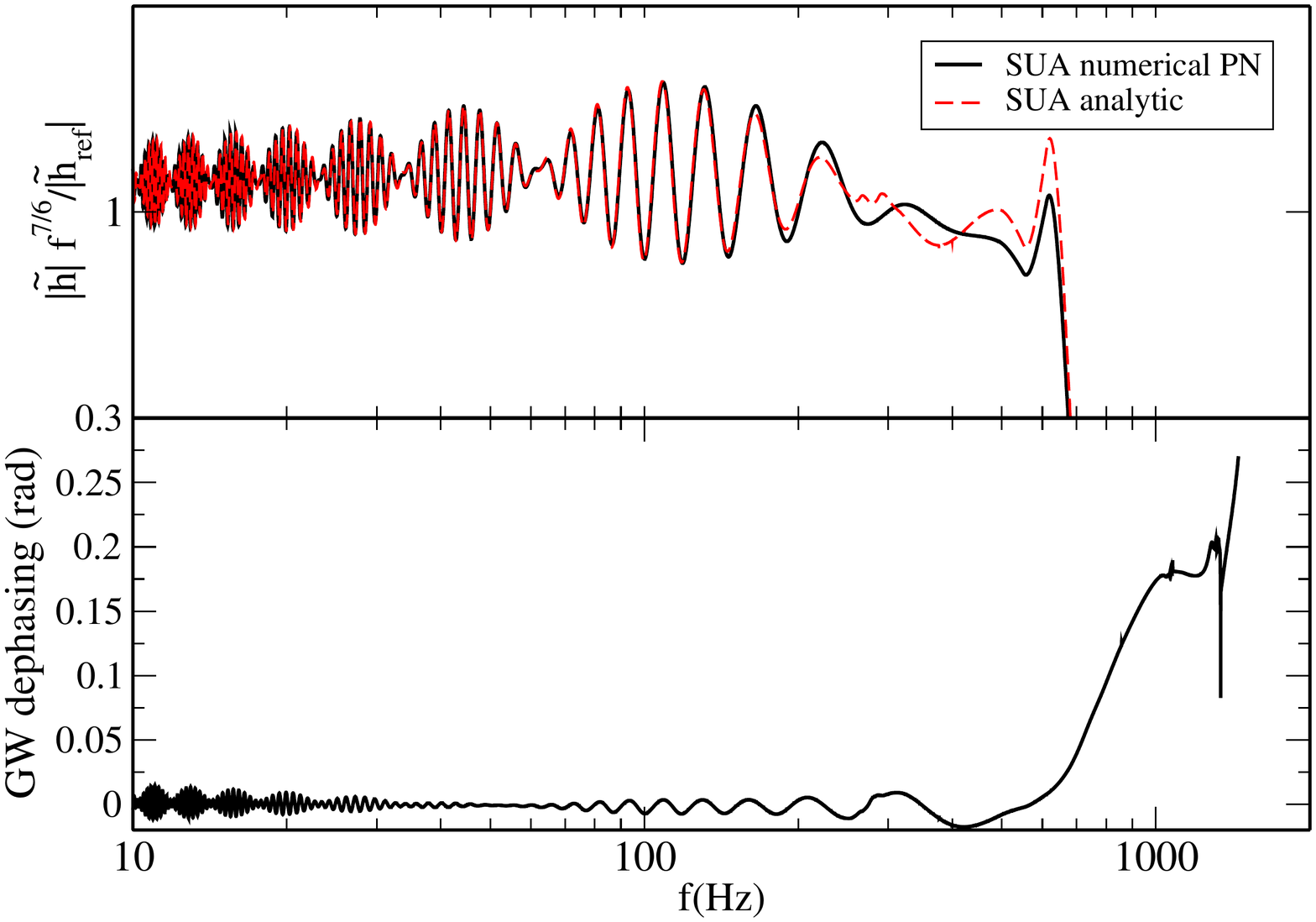}
\includegraphics[width=\columnwidth,clip=true]{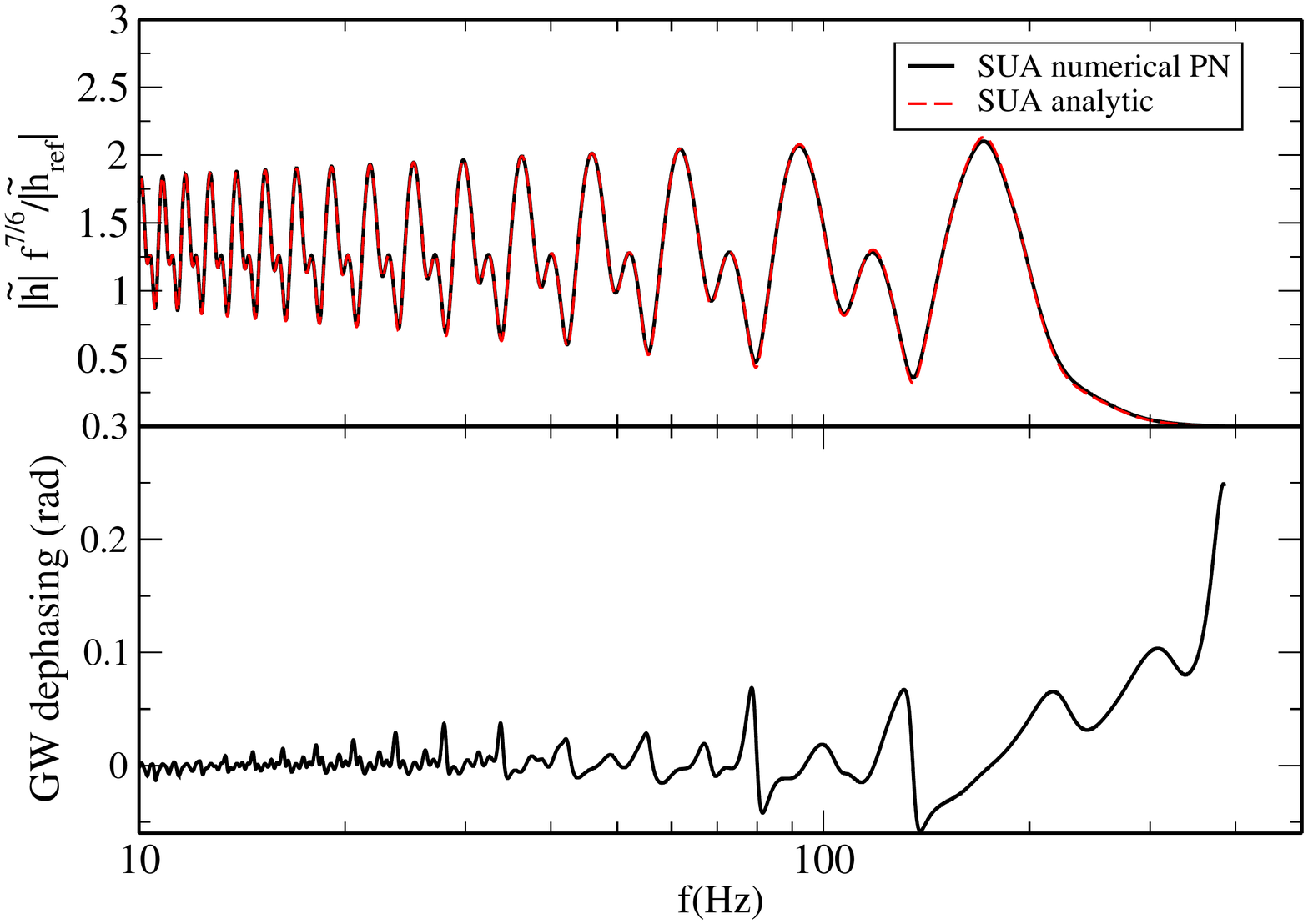}\\
\vspace{-0.5cm}
\includegraphics[width=\columnwidth,clip=true]{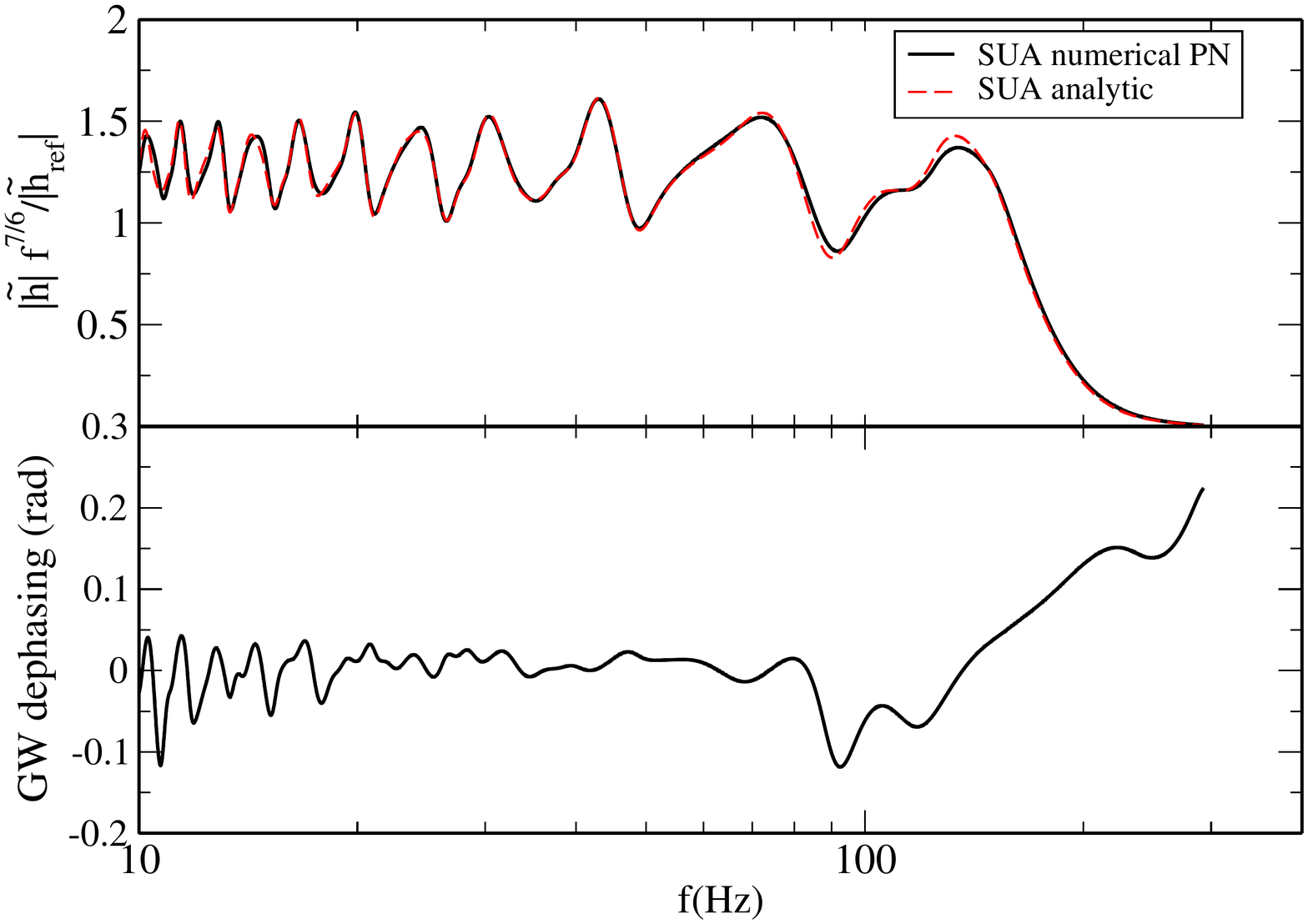}
\includegraphics[width=\columnwidth,clip=true]{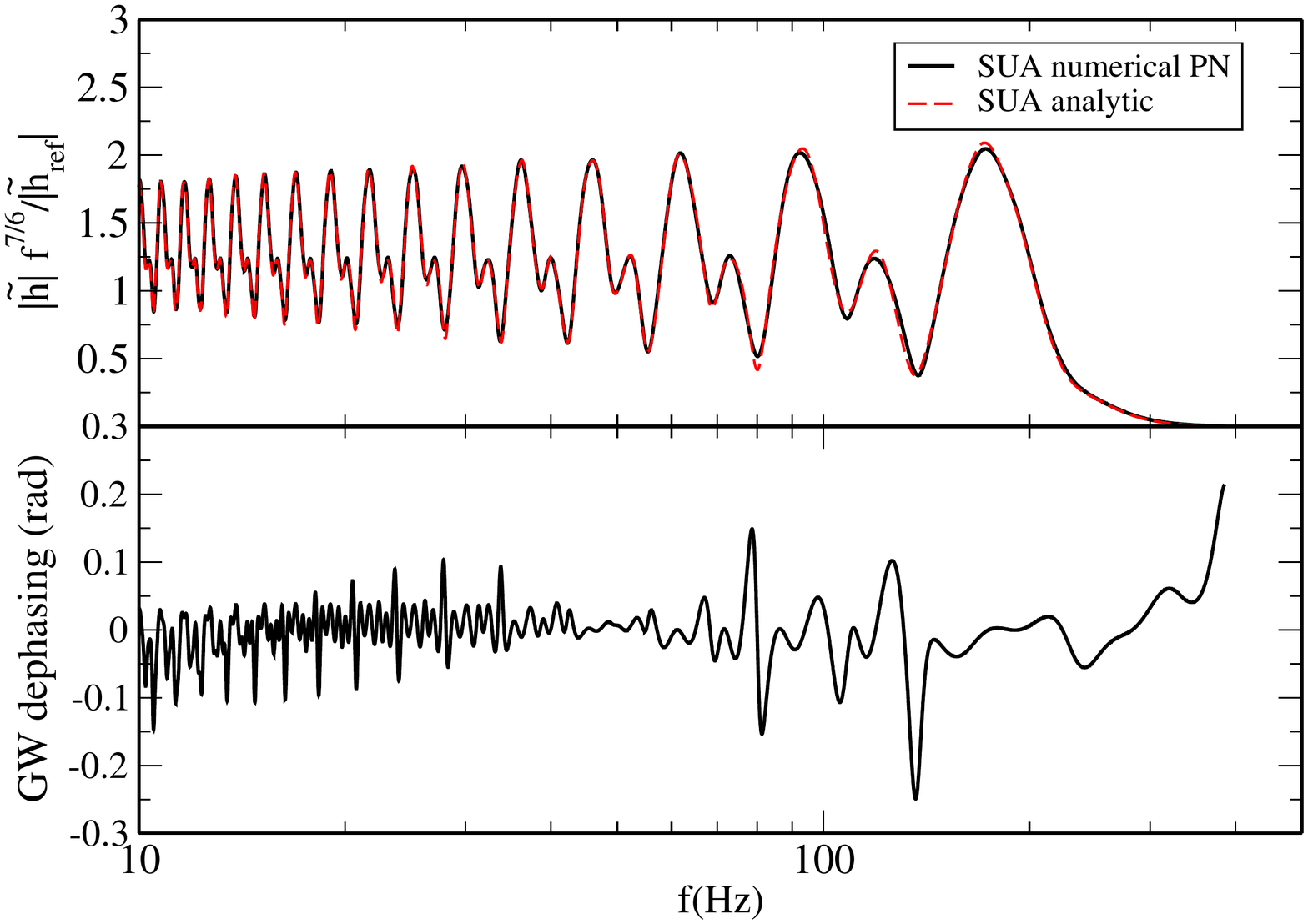}
\vspace{-0.5cm}
\caption{\label{fig:GWcomparisons} (Top Panel) Amplitude of the GW including only the dominant $(\ell=2,m=2)$ harmonic as a function of the GW frequency for the numerical PN and analytic SUA waveforms  for the NSNS (Top Left), the BHNS (Top Right), the BHBH (Bottom Left), and the HSNSBH (Bottom Right) system of Table~\ref{table:systems}. The reference amplitude $|\tilde{h}_{\rm{ref}}|$ is the numerical PN SUA amplitude at 50Hz. (Bottom Panel) GW dephasing between the numerical PN and analytic SUA waveforms.}
\end{center}
\end{figure*}

The polarization states can be decomposed into a spin-weighted spherical 
harmonic basis~\cite{Arun:2008kb,PhysRevD.84.124011,Lundgren:2013jla}
\begin{align}
 h_+ - i h_\times &= \sum_{l \geq 2} \sum_{m = -l}^l H^{lm}(
\theta_s, \phi_s) e^{- i m \Phi},
\end{align}
where
\be
\Phi  =  \phi\sub{orb} - 3  v^3 (2 - \eta  v^2 ) \ln v\,,
\ee 
and $(\theta_s,\phi_s)$ are 
the spherical angles of $\uvec{N}$ in a frame where $\uvec{J}$ is aligned with
the $z$-axis, $\phi\sub{orb}$ is the orbital phase, and
\begin{align}
 H^{lm} &= h^{lm} \sum_{m' = -l}^l D^l_{m',m} (\phi_z, \theta_L, \zeta)
{}_{-2}Y_{lm'} (\theta_s, \phi_s),
\end{align}
where
${}_s Y_{lm}$ are the spin-weighted spherical harmonics, the amplitudes $h^{lm}$ are in~\cite{Blanchet:2014}, $D^l_{m,m'}$ are 
the Wigner 
D-matrices, the angles 
$\theta_L$ and $\phi_z$ are the spherical angles of $\uvec{L}$ in 
the frame where $\uvec{J}$ is aligned with
the $z$-axis, and $\zeta$ satisfies 
$\dot{\zeta} =  \dot{\phi_z} 
\cos{\theta_L}$. In order to solve for $\zeta$ we can employ the same techniques as for $\phi_z$, namely MSA. An explicit expression for $\zeta$ is given in Appendix~\ref{coeff-zeta}.

The above prescribe a waveform $h(t)$ in the time domain. To compute its Fourier 
transform, we use the shifted uniform asymptotics method of~\cite{Klein:2014bua}
\begin{align}
 \tilde{h}(f) &= \sqrt{2\pi} \sum_{m \geq 1} 
 T_m e^{i(2 \pi f t_m - m \Phi - \pi/4)} \nonumber\\
&\times \sum_{l 
\geq 2} \sum_{k=-k\sub{max}}^{k\sub{max}} \frac{a_{k,k\sub{max}}}{2 - 
\delta_{k,0}} \mathcal{H}_{lm}( t_m + k T_m )\label{SUAGW},
\end{align}
where $t_m$ and $T_m$ are defined by
\begin{align}
 2 \pi f &= m \dot{\Phi}(t_m), \label{statcond} \\
 T_m &= \frac{1}{\sqrt{m \ddot{\Phi}(t_m)}}, \\
 \mathcal{H}_{lm} &= \frac{1}{2} (F_+ + i F_\times) \nonumber\\
 &\times \sum_{m' = -l}^l  h^{lm}  D^l_{m',m} (\phi_z, \theta_L, \zeta)
{}_{-2}Y_{lm'} (\theta_s, \phi_s) \nonumber\\
&+ \frac{1}{2} (F_+ - i F_\times) \nonumber\\
 &\times \sum_{m' = -l}^l  h^{l,-m}  D^l_{m',-m} (\phi_z, \theta_L, 
\zeta)
{}_{-2}Y_{lm'} (\theta_s, \phi_s),
\end{align}
and the constants $a_{k, k\sub{max}}$ satisfy the linear system
\begin{align}
 \frac{(-i)^p}{2^p p!} &= \sum_{k=0}^{k\sub{max}} a_{k, k\sub{max}} 
\frac{k^{2p}}{(2p)!},
\end{align}
for $p \in \{ 0, \ldots, k\sub{max} \}$.
In this expression, Eq.~\eqref{statcond} expresses the 
stationary time $t_m$ as a function of the frequency $f$. For a LIGO-type detector, $\mathcal{H}_{lm}$ depends on time 
through $\phi_z$, $\theta_L$, and $\zeta$.

Figure~\ref{fig:GWcomparisons} compares the frequency domain GWs for the $4$ systems of Table~\ref{table:systems} using only the leading $(\ell=2,m=2)$ harmonics of Eq.~\eqref{SUAGW}. The two waveforms are computed with the numerical solution to the PN precession equations and with the analytic solution described in Secs.~\ref{precession} and~\ref{radiationreaction}. Both waveforms are Fourier-transformed with SUA, allowing us to assess the effect of our new analytic solution to the GW amplitude and phase. The agreement between the wave amplitudes is excellent over a wide range of frequencies, while the dephasing between the two waveforms never exceeds $0.3$ radians, even for our BHBH system. This figure serves as a first indication of the accuracy of our model to accurately capture generic precessing features in GWs.

%%%%%%%%%%%%%%%%%%%%%%%%%%%%%%%%%%%%%%%%%%%%%%%%%%%%%%%%%%
\section{Waveform Comparison}
\label{sec:match}

In order to have a more complete picture of our waveform's ability to model 
generic systems, we carry out a Monte Carlo study 
randomizing over the $15$ parameters describing a quasicircular compact binary 
waveform. For the randomization, we draw the components' masses form a flat 
distribution in log space between $[1,2.5]M_{\odot}$ for NSs and 
$[2.5,20]M_{\odot}$ for BHs, while the components' spin magnitudes are uniformly 
distributed in $[0,0.1]$ for NSs and $[0,1]$ for BHs. We selected seemingly low 
BH masses in order to focus on systems for which the inspiral part is the 
most important. Indeed, those are the ones for which the accurate modeling of 
the precession effects are the most challenging, due to the increased number of 
precession cycles that low masses entail. All directions (spin, sky location, 
orbital angular momentum) are drawn uniformly on a unit sphere. The phase of 
coalescence is assumed to be uniform in $[0,2\pi]$, while the time of 
coalescence and the distance are fixed at $10^5$ seconds and $100$Mpc 
respectively.

The large number of systems simulated can only be analyzed through some appropriate and efficient statistic; we use the~\emph{faithfulness} (or~\emph{match}) defined as
\begin{align}
F &\equiv\underset{ t_c, \phi_c}{\text{max }} 
\frac{\left(h_{1}\left|\right.h_{2}\right)}{\sqrt{\left(h_{1}\left|\right.h_{1}\right)
\left(h_{2}\left|\right.h_{2}\right)}}.
\label{match}
\end{align}
The faithfulness is calculated between two waveforms $h_1$ and $h_2$ with the same physical parameters, but maximized over any unphysical parameters: the time $t_c$ and phase $\phi_c$ of coalescence. As such, it is a good estimator of a model's suitability for parameter estimation. The faithfulness always falls between $-1$ and $1$, with the latter indicating perfect agreement between the waveforms. 

Unlike \emph{fitting factors}\footnote{A fitting factor is the faithfulness maximized over all model parameters, quantifying a model's suitability for detection.}, selecting a value for the faithfulness that is ``good enough'' is not straightforward. The nominal fitting factor threshold of $0.965$ corresponds to a $10\%$ drop in detection rates. On the other hand, a faithfulness threshold should be translatable to a requirement about parameter estimation accuracy: the systematic mismodeling error should be smaller than the statistical measurement error. The latter depends on the signal-to-noise ratio (SNR) of the signal, while the former does not, meaning that any faithfulness threshold should take the strength of the signal into account. In Appendix~\ref{f} we calculate the faithfulness threshold as a function of the SNR and find that for an SNR of 10(25)[50], a faithfulness of 0.96(0.9936)[0.9984] suffices for accurate parameter estimation. Led by the SNR of the first detected GW~\cite{Abbott:2016blz}, we set our faithfulness threshold to $0.994$.

In our study $h_1$ is a waveform calculated by numerically solving the precession equations, while $h_2$ uses our new analytic solution. Both waveforms are Fourier-transformed with the SUA method, justified by~\cite{Klein:2014bua} where it was shown that SUA induces a negligible loss of faithfulness compared to a discrete Fourier transform. The use of SUA in both waveforms allows us to isolate the effect of our new solution: any mismatch is solely caused by the solution to the precession equations described in this paper.

The inner product in Eq.~\eqref{match} is defined in the usual way
\begin{align}
 \left(h_{1}\left|\right.h_{2}\right) &\equiv 
4 \Re \int_{f_{\min}}^{f_{\max}} \frac{\tilde{h}_1(f) \tilde{h}_2^*(f)}{S_{n}(f)} \; 
df\,,
\end{align}
where $f_{\min}=10$Hz is aLIGO's lower frequency cutoff, $f_{\max}$ is the frequency that corresponds to an orbital separation of $6M$, and $S_{n}(f)$ is aLIGO's design zero-detuning, high power noise spectral density~\cite{AdvLIGO-noise}. 

\begin{figure*}[htbp]
\begin{center}
\includegraphics[width=\columnwidth,clip=true]{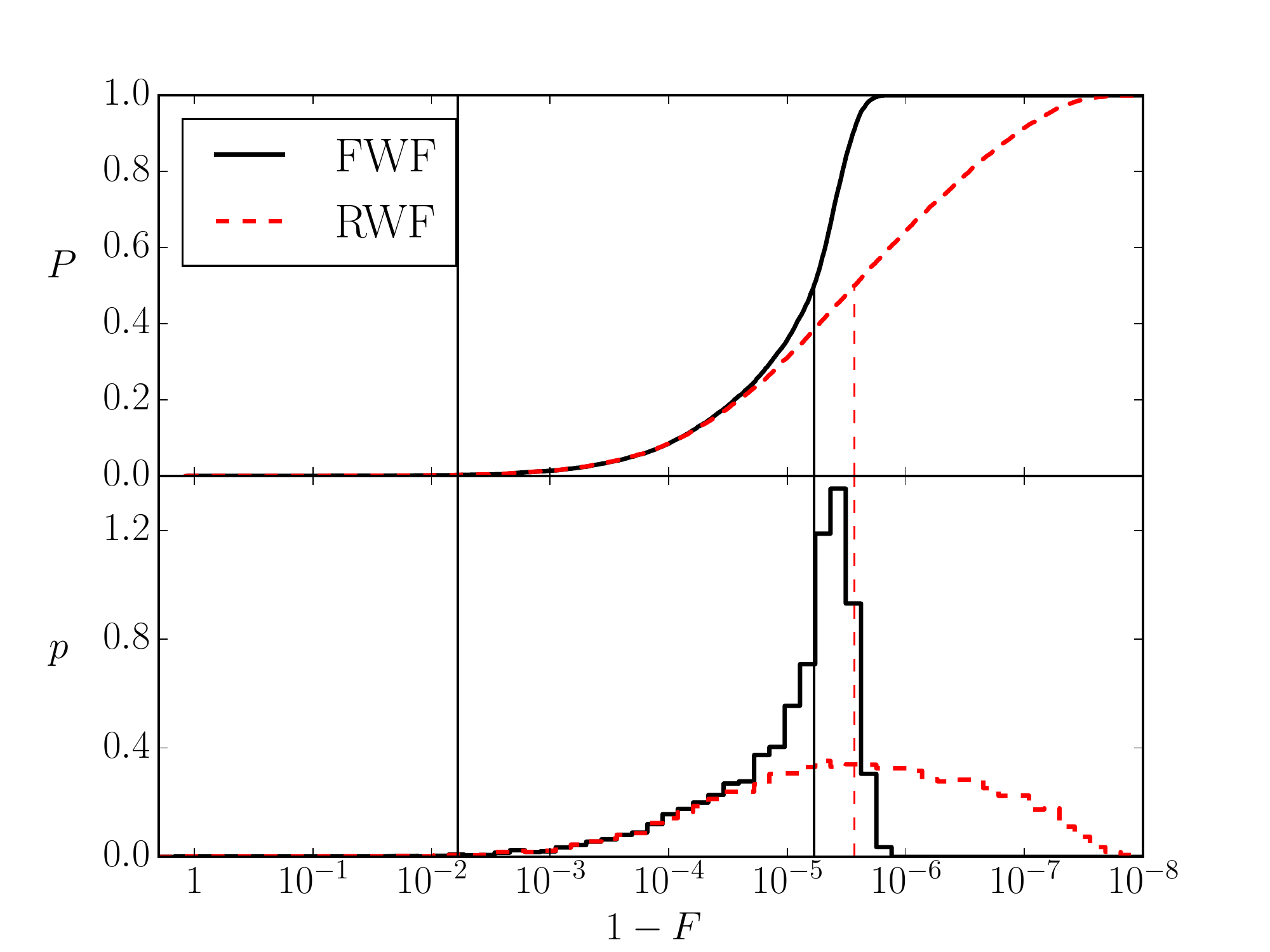}
\includegraphics[width=\columnwidth,clip=true]{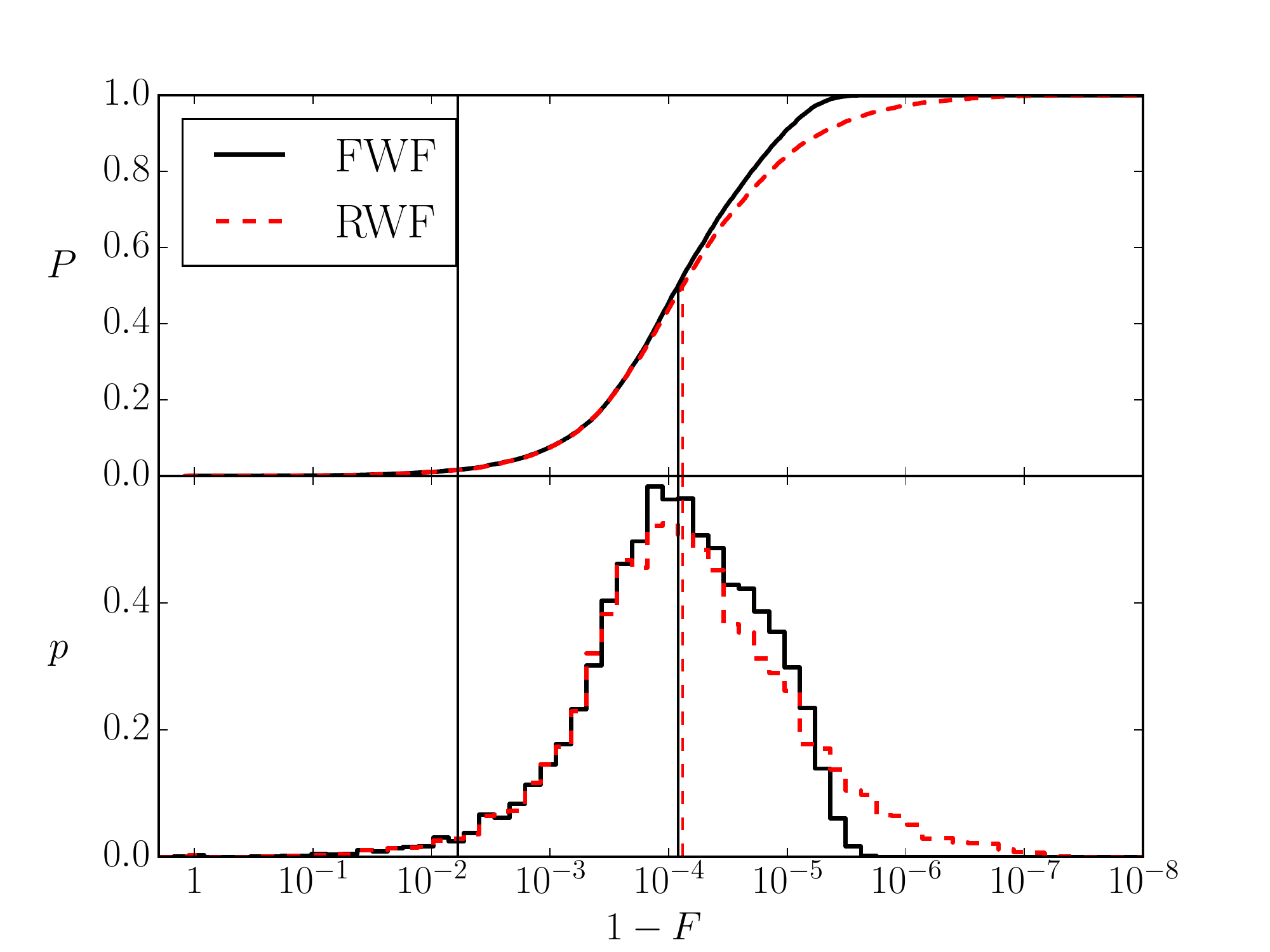}\\
\includegraphics[width=\columnwidth,clip=true]{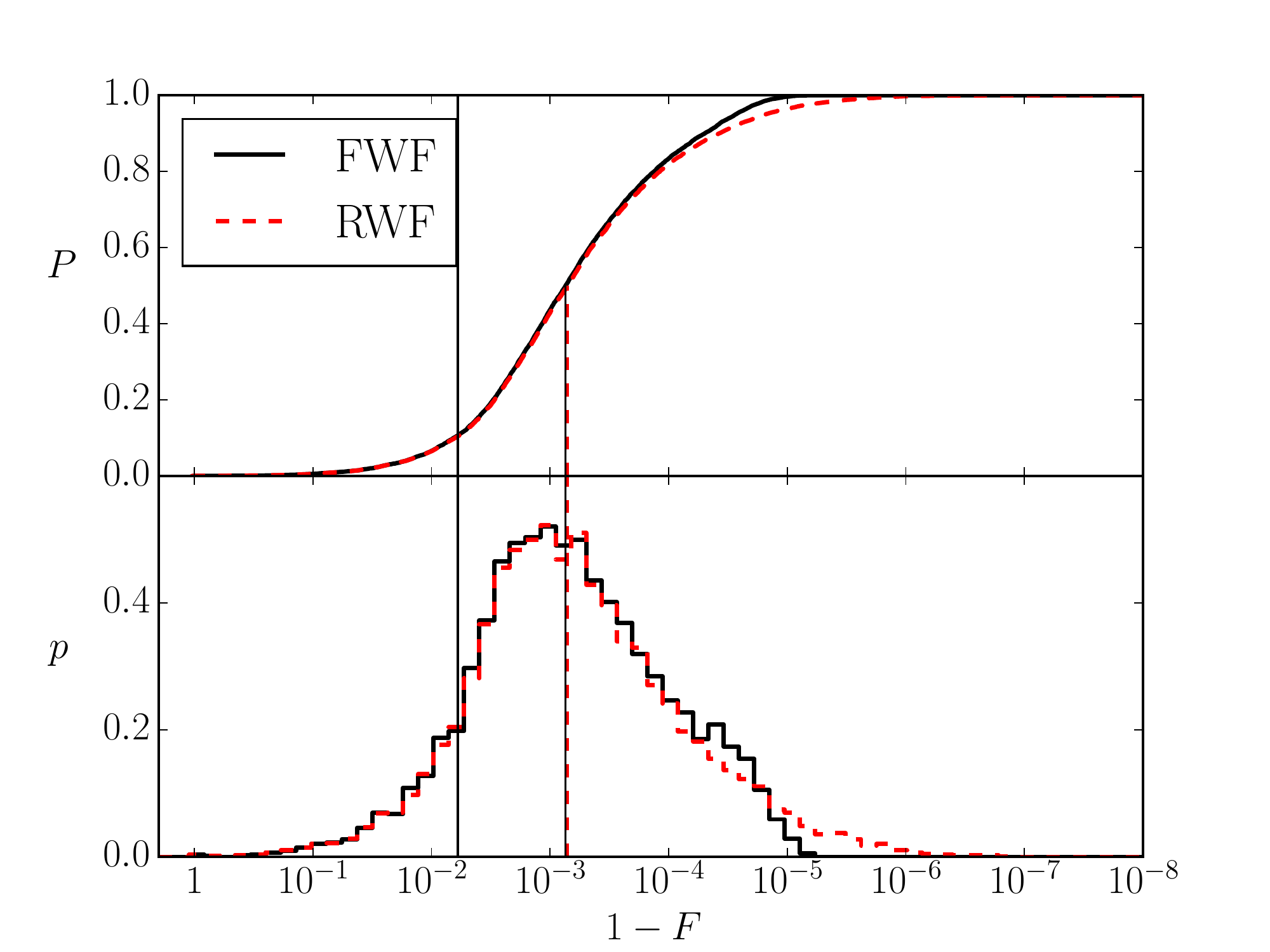}
\includegraphics[width=\columnwidth,clip=true]{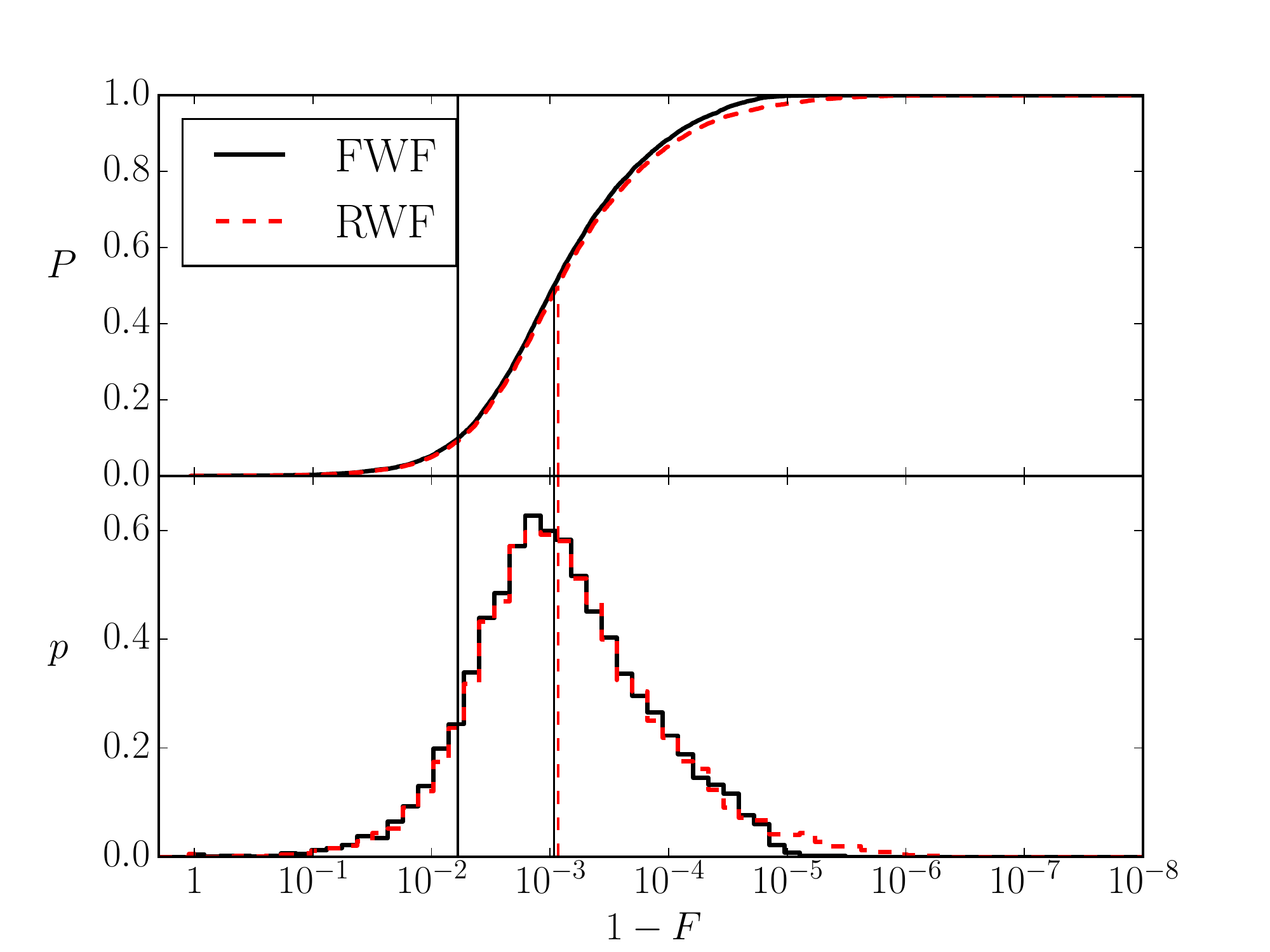}
\caption{\label{fig:matches} 
Distribution of $1-F$ for 
NSNS (Top Left), BHNS (Top Right), BHBH (Bottom Left) and the $4\text{th}$ generic set containing all masses and spins (Bottom Right) for waveforms with full harmonic 
content (solid black line) and waveforms restricted to the leading $(\ell=2, 
m=2)$ mode (dashed red line). Top panels show cumulative distribution functions, and bottom panels give the corresponding probability distribution function. The leftmost 
vertical line denotes a faithfulness $F = 0.994$, corresponding to $0.3\%$ of 
systems for NSNS (both waveforms), $1.6\%$ for BHNS (both waveforms), $10.4\%$ 
(RWF) and $10.7\%$ (FWF) for BHBH, and $9.2\%$ (RWF) and $9.9\%$ (FWF) for 
the $4\text{th}$ set. The other vertical lines correspond to the medians 
of the distributions, which are $1-F = 2.7 \times 10^{-6}$ (RWF), and $1-F = 6 \times 10^{-6}$ (FWF) for NSNS, $1-F = 7.6 \times 10^{-5}$ (RWF), and $1-F = 8.3 \times 10^{-5}$ (FWF) for NSBH, $1-F = 7.1 \times 10^{-4}$ (RWF), and $1-F = 7.4 \times 10^{-4}$ (FWF) for BHBH, and $1-F = 9.3 \times 10^{-4}$ (RWF), and $1-F = 8.6 \times 10^{-4}$ (FWF) for the $4\text{th}$ set.}
\end{center}
\end{figure*}

Figure~\ref{fig:matches} shows the distributions
of $1-F$ for $4$ sets, each containing $10,000$ systems. The first $3$ sets contain systems with masses and spins corresponding to NSNS, BHNS, and BHBH systems respectively. The fourth set contains an additional, less astrophysically motivated but useful to test our model in the most challenging setting, type of system where both masses were drawn from a log-flat distribution 
ranging from $1M_\odot$ to $20 M_\odot$, and both spin magnitudes 
uniformly distributed in [0, 1]. We study $2$ 
different types of waveforms:~\emph{full} waveforms (FWF) contain all the known 
harmonics in Eq.~\eqref{SUAGW}, while~\emph{restricted} waveforms (RWF) contain only 
the dominant $(\ell=2, m=2)$ harmonic. 

\begin{table}
\begin{centering}
\begin{tabular}{ccccccccccc}
\hline
\hline
\noalign{\smallskip}
Waveform  &&  Threshold   &&  \multicolumn{1}{c}{NSNS} &&  \multicolumn{1}{c}{BHNS} &&  \multicolumn{1}{c}{BHBH} &&  \multicolumn{1}{c}{HSNSBH}  \\
\hline
\noalign{\smallskip}
RWF   &&  $0.965$ &&$0.06\%$ && $0.33\%$   && $1.85\%$ && $1.14\%$ \\
RWF   && $0.994$ && $0.3\%$ && $1.6\%$  && $10.4\%$ && $9.2\%$\\
FWF   && $0.965$ && $0.06\%$ && $0.33\%$   && $1.85\%$ && $1.26\%$ \\
FWF   && $0.994$ &&$0.3\%$ && $1.6\%$  && $10.7\%$ && $9.9\%$\\
\noalign{\smallskip}
\hline
\hline
\end{tabular}
\end{centering}
\caption{Percentages of subthreshold systems encountered in our analysis for each type of system.}
\label{table:Mresults}
\end{table}

The agreement between our analytical waveform and the numerical PN one is 
excellent for a wide range of parameters. In the NSNS case we find that only 
$0.06\%(0.3\%)$ of the systems have a faithfulness below the $0.965(0.994)$ for 
both waveforms, while for BHNS systems, this number is $0.33\%(1.6\%)$ for both 
waveforms. The percentage of systems below the nominal faithfulness threshold is 
increased to $1.85\%(10.4\%)$ (RWF) and $1.85\%(10.7\%)$ (FWF) in the case of 
BHBHs and $1.14\%(9.2\%)$ (RWF) and $1.26\%(9.9\%)$ (FWF) for the 
$4\text{th}$ generic set. This increase is not unexpected, since precessional 
features are more pronounced, and hence more difficult to model, when the spins 
are large and the masses different. In the next section we study the various 
sources of error in our analytical waveform and quantify their effect. Table~\ref{table:Mresults}
summarizes these results.

%%%%%%%%%%%%%%%%%%%%%%%%%%%%%%%%%%%%%%%%%%%%%%%%%%%%%%%%%%
\section{Source of Error}
\label{error}

The subthreshold systems of Fig.~\ref{fig:matches} can be split into two rough categories: systems for which the faithfulness is very low $F\lesssim 0.8$, and systems for which the faithfulness is high, but not high enough $0.8\lesssim F \lesssim 0.994$. Systems falling into the first category can mainly be explained by the effect described in Appendix~\ref{phizexpressions}. For them the orbital angular momentum becomes approximately (anti)aligned with the total spin angular momentum at some point in the evolution of the systems. In this case the PN expansion of Eq.~\eqref{phizdotfinal} becomes ill-defined. Our specific choice for the values of $D_2$ and $D_4$ in Eqs.~\eqref{D2-def} and~\eqref{D4-def} to some extent ameliorates this problem, yet it does not fully solve it. We have explored many choices for $D_2$ and $D_4$, some even leading up to $8\%$ of systems with faithfulnesses below $0.96$ in the BHBH case. The particular values for $D_2$ and $D_4$ we employ in our model [Eqs.~\eqref{D2-value} and~\eqref{D4-value}] yield the best results among all the expressions we tested. 

Systems falling in the second category can be modeled accurately only for low SNR signals. The unfaithfulness of these systems can be attributed to the various approximations we have used in our model construction. In order to quantify the effect of each approximation, we retrace the steps we followed in Sec.~\ref{radiationreaction} to add radiation reaction effects to the exact precession solution of Sec~\ref{precession}.

\begin{enumerate}
\item Our first task when adding radiation reaction effects is to specify a coordinate system. In Sec.~\ref{frame-rr} we assume that $\uvec{J}$ is constant and identify it with the $\uvec{z}$ axis of our system.
\item In Sec.~\ref{Jmag} we use MSA to solve for the magnitude of the total angular momentum $J$.
\item In Sec.~\ref{Smag} we use MSA and a PN approximation to solve for the total spin magnitude $S$.
\item Finally, in Sec~\ref{phizsol} we use the $J$ and $S$ obtained above to solve for the precession angle $\phi_z$.
\end{enumerate}

Overall, the addition of radiation reaction effects requires the identification of a coordinate system and the solution to $3$ coupled differential equations. Below we perform each of these steps numerically and each time compute matches for BHBH systems in order to quantify the improvement. The unfaithfulness distributions are given in Fig.~\ref{fig:matchesstudy}, while the inset focuses on the region of interest $F\in[0.9,0.999]$ with the vertical line denoting $F=0.994$. The different curves in this figure represent the following:
\begin{figure}[htbp]
\begin{center}
\includegraphics[width=\columnwidth,clip=true]{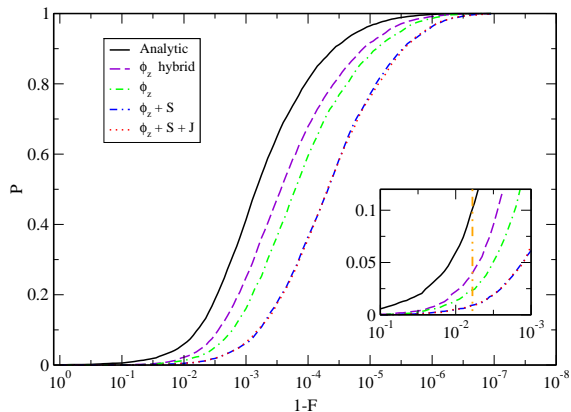}
\caption{\label{fig:matchesstudy} 
Cumulative distribution of $1-F$ for waveforms including different degrees of numerical and analytical calculations. The difference between these distributions is a quantification of the error from each analytic approximation we have made. The inset shows the faithfulness range of interest $F\in[0.9,0.999]$, while the vertical line denotes the faithfulness threshold $F=0.994$. See the text for more details and discussion.}
\end{center}
\end{figure}
\begin{itemize}
\item
The black solid line corresponds to the fully analytic waveform also studied in Fig.~\ref{fig:matches}, with approximately $10\%$ of the systems below $0.994$. 
\item
The maroon dashed line was created with a waveform that evaluates Eq.~\eqref{phiz-full} for $\phi_z$ with the analytical $J$ and $S$, and then solves this equation numerically, bringing the percentage of subthreshold systems down to $\sim4\%$. 
\item
The green dot-dashed line again uses a numerical solution to Eq.~\eqref{phiz-full} but where now $S$ and $J$ are obtained and substituted by numerically solving their corresponding differential equations. We stress that in this version of the waveform, $S$ and $J$ are solved for numerically \emph{only} when used in Eq.~\eqref{phiz-full}. This corresponds to the most accurate solution for $\phi_z$ possible and results in $\sim2\%$ of subthreshold systems.
\item
The blue dot-double-dashed line again uses the most accurate $\phi_z$ from the previous waveform, and now the numerical $S$ is also used for the entire waveform. This brings the percentage of subthreshold systems down to $\sim 1\%$.
\item
Finally, the red dotted line is produced with a waveform that solves for all $\phi_z$, $S$, and $J$ numerically. The faithfulness distribution for this waveform is almost indistinguishable from the previous waveform with $\sim 1\%$ of subthreshold systems.
\end{itemize}

The above results suggest a clear-cut way to improve our model if a more faithful waveform model is required in the future. The largest improvement would be obtained if we found a more accurate solution to Eq.~\eqref{phiz-full} for the precession angle $\phi_z$, either by improving the PN solution to Eq.~\eqref{phizPN}, or by taking Eq.~\eqref{phiz0} to higher order in MSA. We have studied those two error sources separately, and have found the former to dominate over the latter. The second step would be to improve the solution for the total spin magnitude by employing elements of MSA to solve Eq.~\eqref{dSdt-final}, or a more accurate PN prescription for Eq.~\eqref{psidot-rr} for the phase of $S$. Finally, improving the solution for the magnitude of the total angular momentum $J$ will not affect the waveform considerably.

Even with these improvements, we still have $\sim1\%$ of our systems with $F<0.994$. The only approximation we have made for these systems is that the total angular momentum has a fixed direction, and to verify that this approximation indeed breaks, we studied $100$ of these systems explicitly. We evolved $\vec{J}$ and found that $2$ of these systems undergo the effect of transitional precession~\cite{Apostolatos:1994mx}. The remaining $98$ systems do not undergo transitional precession but (i) have parameters that are statistically consistent with what is required for transitional precession (small mass ratios, large and misaligned spin for the larger body, everything else random), and (ii) the components $J_x$ and $J_y$ always remain smaller than $J_z$, but by about an order of magnitude only. Compare this with Fig.~\ref{fig:Jcomparisons} where $J_x$ and $J_y$ remain at least $3$ orders of magnitude below $J_z$ for all frequencies.

As a concluding remark, we should mention the effect of the inclination angle $\bm{L}\cdot\bm{N}$ on our results. It is well known that binaries observed approximately edge-on ($\bm{L}\cdot\bm{N}\sim0$) exhibit the largest precessional effects. We indeed find that all other things being equal, the more edge-on the binary is, the lower its faithfulness. This is because when precessional effects become more pronounced, a better and better modeling of them is required in order to achieve a certain goodness of fit. In other words, inclination does not cause problems on its own, but rather it amplifies preexisting ones, an effect also observed in~\cite{Abbott:2016wiq}. We should note, however, that edge-on systems are less likely to be detected by aLIGO due to selection effects~\cite{Littenberg:2015tpa}.

%%%%%%%%%%%%%%%%%%%%%%%%%%%%%%%%%%%%%%%%%%%%%%%%%%%%%%%%%%
\section{Discussion}
\label{conclusion}

We have constructed the first closed-form fully analytic GW template in the frequency domain that can accurately model quasicircular systems of generic masses, spin magnitudes, and spin orientations in the inspiral phase. We expand the exact solution to the precession equations in the absence of radiation reaction derived by Kesden et al.~\cite{Kesden:2014sla} to include radiation reaction using elements from \emph{multiple scale analysis}. This allows us to derive the first closed-form time-domain GW model valid for generic inspirals. We then use the method of \emph{shifted uniform asymptotics} to transform this waveform from the time domain to the frequency domain.

The resulting waveform is ideal for extracting parameters from generically precessing quasicircular inspirals as demonstrated by a Monte Carlo study of $40,000$ system; only $0.8\%(5.4\%)$ of them had a faithfulness with a numerical PN waveform that solves the precession equations numerically below $0.965(0.994)$. The remaining inaccuracies of our model can be mapped back to specific assumptions we made while solving the spin-precession equations including radiation reaction. Analytical understanding of all these assumptions and the elements that enter our waveform construction enable us to improve the accuracy of our model if deemed necessary when more sensitive GW detector networks become available. This is, perhaps, the most attractive feature of having analytic control over complicated processes like spin-precession.

Finally, analytic methods have the potential to be much faster than numerical ones, while still encompassing all precessional effects. We estimate that our analytic SUA waveform can be up to 15 times faster to evaluate than the numerical SUA waveform in certain regions of the parameter space. Interestingly, the region of the parameter space where the analytic SUA waveform presents the maximum improvement over the numerical SUA waveforms (the BHBH case) is distinct from the region where the numerical SUA waveform is much faster than fully numerical PN time-domain models (the NSNS case)~\cite{Klein:2014bua}. This suggests that a hybrid model where the numerical or the analytical SUA is called depending on the system's mass can achieve both high accuracy and numerical efficiency. Further improvement could be obtained through reduced order modeling and reduced order quadrature integration in data analysis implementations~\cite{Smith:2016qas}. We leave such studies to future work.
 
As a final remark, we note that our results lay the framework for the construction of full inspiral-merger-ringdown (IMR) waveforms following similar procedures as IMRPhenomP~\cite{Hannam:2013oca,Schmidt:2012rh,Schmidt:2014iyl}. This is a promising avenue for future research since such an IMR waveform has the potential to be more accurate than the IMRPhenomP due to more accurate description of precessional dynamics, as well as faster than SEOBNRv3~\cite{Pan:2013rra} due to the analytic treatment of the inspiral dynamics.

%%%%%%%%%%%%%%%%%%%%%%%%%%%%%
\acknowledgments

We would like to thank Emanuele Berti, Mike Kesden, Sylvain Marsat, and Frank Ohme for helpful discussions.
K.~C. acknowledges support from the Onassis Foundation. N.~Y. acknowledges 
support NSF CAREER Grant No. PHY-1250636. N.~C. 
acknowledges support from the NSF Award PHY-1306702.
N.~C. and N.~Y. acknowledge support from NASA Grant No. 
NNX16AB98G. A.~K. is supported by NSF CAREER Grant No. PHY-1055103, by FCT 
contract IF/00797/2014/CP1214/CT0012 under the IF2014 
Programme, and by
the H2020-MSCA-RISE-2015 Grant No. StronGrHEP-690904. This work was supported by 
the Centre National d'{\'E}tudes 
Spatiales.

%%%%%%%%%%%%%%%%%%%%%%%%%%%%
\appendix

%---------------------------------------------------------------------------
\section{Coefficients of $\dot{ v}$}
\label{vdot-coeff}

The coefficients of the  evolution of the PN parameter $ v$ as defined in Eq.~\eqref{vdot} are
\begin{align}
 g_0 & = \frac{1}{a_0},
\\
 g_2 & = -\frac{a_2}{a_0},
\\
 g_3 & = -\frac{a_3}{a_0},
\\
 g_4 & = -\frac{a_4-a_2^2}{a_0},
\\
 g_5 & = -\frac{a_5-2a_3a_2}{a_0},
\\
 g_6 & = -\frac{a_6-2a_4a_2-a_3^2+a_2^3}{a_0},
\\
 g^{\ell}_6 & = -\frac{3b_6}{a_0},
\\
 g_7 & = -\frac{a_7-2a_5a_2-2a_4a_3+3a_3a_2^2}{a_0},
\end{align}
with all other terms vanishing. The coefficients $\{a_{i},b_{i}\}$ are given in Appendix A of~\cite{Chatziioannou:2013dza}. The spin couplings in the above expressions are evaluated with all angular momenta averaged over one precession cycle using the solution of Secs.~\ref{precession} and~\ref{radiationreaction}. The error induced by this is of 4PN order, higher that the order to which we know the $\dot{ v}$ expansion. Explicitly, we use
\begin{align}
&\bm{S}_1 \cdot \uvec{L} \rightarrow \precav{\bm{S}_1 \cdot \uvec{L}}=\frac{c_1(1+q)-q\eta\xi}{\eta(1-q^2)},
\\
&\bm{S}_2 \cdot \uvec{L} \rightarrow  \precav{\bm{S}_2 \cdot \uvec{L}}=-q\frac{c_1(1+q)-\eta\xi}{\eta(1-q^2)},
\\
&\bm{S}_1 \cdot \bm{S}_2 \rightarrow \precav{\bm{S}_1 \cdot \bm{S}_2}= \frac{S^2_{\av}}{2}-\frac{S_1^2+S_2^2}{2},
\\
&(\bm{S}_1 \!\cdot \!\uvec{L})^2 \rightarrow \precav{(\bm{S}_1\! \cdot \!\uvec{L})^2}\!=\!\precav{\bm{S}_1\! \cdot \!\uvec{L}}^2 \!+\! \frac{(S_+^2-S_-^2)^2 v_0^2}{32\eta^2(1-q)^2},
\\
&(\bm{S}_2 \!\cdot \!\uvec{L})^2 \rightarrow  \precav{(\bm{S}_2\! \cdot\! \uvec{L})^2}\!=\!\precav{\bm{S}_2\! \cdot \!\uvec{L}}^2\! +\! \frac{q^2(S_+^2-S_-^2)^2 v_0^2}{32\eta^2(1-q)^2},
\\
&(\bm{S}_1 \cdot \uvec{L})(\bm{S}_2 \cdot \uvec{L}) \rightarrow  \precav{(\bm{S}_1 \cdot \uvec{L})(\bm{S}_2 \cdot \uvec{L})}\nn
\\
&\quad=\precav{\bm{S}_1 \cdot \uvec{L}}\precav{\bm{S}_2 \cdot \uvec{L}} - \frac{q(S_+^2-S_-^2)^2 v_0^2}{32\eta^2(1-q)^2},
\end{align}
where $ v_0$ corresponds to the value of $ v$ at the initial time, $ \xi, S^2_{\av}, c_1$ are defined in Eqs.~\eqref{xi-def},~\eqref{Save-def}, and~\eqref{Jsol} respectively, and $S_+^2,S_-^2$ are the roots of the right hand side of Eq.~\eqref{dSdt-prel}.

%---------------------------------------------------------------------------
\section{Coefficients of the precession solution}
\label{coeff}

Below, we provide explicit expressions for the coefficients appearing in the precession solution of Sec.~\ref{precession}. 

The coefficients of Eq.~\eqref{dSdt-prel} are
\begin{align}
A &= -\frac{3}{2\sqrt{\eta}}   v^6\left(1-\xi \,  v\right),
\\
B &= (L^2+S_1^2)q + 2 L \xi - 2 J^2 - S_1^2 - S_2^2+\frac{L^2+S_2^2}{q},
\\
C &= (J^2-L^2)^2 - 2 L \xi (J^2 - L^2) \nn
\\
&- 2\frac{1-q}{q}(S_1^2 - q S_2^2 )L^2+4 \eta  L^2 \xi^2 \nn
\\
&- 2\delta m (S_1^2 - S_2^2) \xi L + 2 \frac{1-q}{q} (q S_1^2 - S_2^2) J^2,
\\
D &\!=\! \frac{1-q}{q}(S_2^2-q S_1^2)(J^2-L^2)^2  +\frac{\delta m^2}{\eta} (S_1^2\! -\! S_2^2)^2 L^2\nn
\\
&+ 2\delta m \,L \xi  (S_1^2 - S_2^2) (J^2-L^2).
\end{align}

The coefficients of Eq.~\eqref{phizdot} are
\begin{align}
a &=  \frac{1}{2} v^6 \left\{ 1+ \frac{3}{2\eta}\left(1-\xi  v\right)\right\},
\\
c_0 &= \frac{3 }{4} (1-\xi  v)  v^2 \left\{ \eta^3 + 4 \eta^3\xi  v \right.\nn
\\
&\left.- 2 \eta \left[ J^2 - S_+^2 + 2 (S_1^2-S_2^2)\delta m\right] v^2   \right.\nn
\\
&\left.- 4 \eta \xi (J^2-S_+^2)  v^3 + \frac{(J^2-S_+^2)^2}{\eta}  v^4\right\},
\\
c_2 &= -\frac{3 \eta }{2 } (S_+^2-S_-^2)\left(1 + 2 \xi  v - \frac{J^2-S_+^2}{\eta^2} v^2\right)\! (1\!-\! \xi v)  v^4,
\\
c_4 &= \frac{3 }{4 \eta}  (S_+^2-S_-^2)^2  (1- \xi v)  v^6,
\\
d_0 &= - [J^2-(L+S_+)^2 ][J^2-(L-S_+)^2],
\\
d_2 &= -2 (S_+^2-S_-^2)(J^2+L^2-S_+^2),
\\
d_4 &= -(S_+^2-S_-^2)^2.
\end{align}

The coefficients of Eq.~\eqref{phiz-norr} are
\begin{align}
A_{\phi} &= A +\frac{c_4}{d_4},
\\
B_{\phi} &= \left( \frac{ c_4}{d_4} - \frac{c_0+c_2+c_4}{d_0+d_2+d_4}\right),
\\
C_{\phi} &= C_1 + C_2,
\\
D_{\phi} &= C_1 - C_2,
\\
n_c &= 2\frac{d_0+d_2+d_4 }{2 d_0+d_2+s_d},
\\
n_d &= \frac{2d_0+d_2+s_d}{2 d_0},
\end{align}
where 
\begin{align}
C_1 & =- \frac{1}{2}\left( \frac{ c_0}{d_0} - \frac{c_0+c_2+c_4}{d_0+d_2+d_4}\right),
\\
C_2 & =  \frac{c_0 (-2 d_0 d_4 + d_2^2 + d_2 d_4) - c_2 d_0 (d_2 + 2 d_4) }{2 d_0 (d_0 + d_2 + d_4) s_d }\nn
\\
&+ \frac{c_4 d_0 (2 d_0 + d_2) }{2 d_0 (d_0 + d_2 + d_4)s_d },
\\
s_d &= \sqrt{d_2^2-4d_0d_4}.
\end{align}
%

%---------------------------------------------------------------------------
\section{Coefficients of $\psi$}
\label{coeff-psi}

The coefficients in Eq.~\eqref{psi_rr} are
\begin{align}
\psi_1 &= 3\frac{2 \xi \eta^2- c_1}{\eta \delta m^2},
\\
\psi_2 &=  \frac{3 g_2}{ g_0} + \frac{3}{2\eta^3}  \left\{2\Delta - 2\frac{\eta^2}{\delta m^2}S^2_{\av}-10\frac{\eta}{\delta m^4}c_1^2\right.\nn
\\
&\left.  + 2\frac{\eta^2}{\delta m^2}\frac{7+6q+7q^2}{(1-q)^2} c_1 \xi -\frac{\eta^3}{\delta m^2}\frac{3+4q+3q^2}{(1-q)^2}\xi^2\right.\nn
\\
&\left.+ \frac{ \eta}{(1-q)^2}\left[q(2+q) S_1^2 + (1+2q) S_2^2\right]\right\},
\end{align}
where
\begin{align}
\Delta&\!=\!\left\{\!
\left\{\!  \frac{c_1^2\eta}{q\delta m^4} -  \frac{2 c_1\eta^3(1+q) }{q\delta m^4}\xi - \frac{\eta^2}{\delta m^4} [\delta m^2 S_1^2 - \eta^2 \xi^2] \!\right\} \right.\nn
\\
&\left. \times \left\{ \! \frac{c_1^2\eta^2}{\delta m^4} - \frac{2 c_1 \eta^3 (1+q)}{\delta m^4}  \xi - \frac{\eta^2}{\delta m^4} [\delta m^2 S_2^2 - \eta^2 \xi^2]\! \right\}\! \right\}^{1/2}\!\!\!.
\end{align}
%

%---------------------------------------------------------------------------
\section{Coefficients of $\phi_z$}
\label{coeff-phiz}

For the PN expansions of the coefficients of Eq.~\eqref{int-phiz} we define
\begin{align}
R_m &= S_+^2 - S_-^2,
\\
c_p &= (S_+^2 \eta^2 - c_1^2),
\\
c_m &= (S_-^2 \eta^2 - c_1^2),
\\
a_1 &= \frac{1}{2}+\frac{3}{4}\eta,
\\
a_2 &= -\frac{3}{4\eta}\xi,
\\
a_d &= \frac{-3 (S_1^2-S_2^2) \eta \,\delta m + 3 \frac{c_1}{\eta} (c_1-2 \xi \eta^2 )}{4\sqrt{c_p c_m}},
\\
c_d &=  \frac{3}{128} \frac{R_m^2}{\eta \sqrt{c_p c_m}} ,
\\
h_d &= \frac{c_1}{\eta^2}\left(1-\frac{c_p+c_m}{2\sqrt{c_pc_m}}\right),
\\
f_d &= \frac{c_p+c_m}{8\eta^4}\left(1-\frac{c_p+c_m}{2\sqrt{c_pc_m}}\right),
\\
\Omega_{z,0} & = a_1 + a_d,
\\
\Omega_{z,1} & = a_2 - a_d \xi - a_d h_d,
\\
\Omega_{z,2} & = a_d h_d \xi + c_d - a_d f_d + a_d h_d^2 ,
\\
\Omega_{z,3} & = (a_d f_d - c_d - a_d h_d^2)(\xi+h_d) + a_d f_d h_d ,
\\
\Omega_{z,4} & = (c_d+ a_d h_d^2-2a_d f_d)(h_d\xi+h_d^2-f_d) - a_d f_d^2 ,
\\
\Omega_{z,5} & = (c_d - a_d f_d + a_d h_d^2)f_d(\xi+2h_d) \nn
\\
&- (c_d+ a_d h_d^2-2a_d f_d)h_d^2(\xi+h_d) -a_df_d^2h_d ,
\end{align}

With these definitions, the coefficients of Eq.~\eqref{int-phiz} are 
\begin{align}
\langle \Omega_z \rangle^{(0)} & = 3  g_0 \Omega_{z,0},
\\
\langle \Omega_z \rangle^{(1)} & = 3  g_0 \Omega_{z,1},
\\
\langle \Omega_z \rangle^{(2)} & = 3 ( g_0 \Omega_{z,2} +  g_2 \Omega_{z,0}),
\\
\langle \Omega_z \rangle^{(3)} & = 3 ( g_0 \Omega_{z,3} +  g_2 \Omega_{z,1} +  g_3 \Omega_{z,0}),
\\
\langle \Omega_z \rangle^{(4)} & = 3 ( g_0 \Omega_{z,4} +  g_2 \Omega_{z,2} +  g_3 \Omega_{z,1} +  g_4 \Omega_{z,0}),
\\
\langle \Omega_z \rangle^{(5)} & = 3 ( g_0 \Omega_{z,5} +  g_2 \Omega_{z,3} +  g_3 \Omega_{z,2} +  g_4 \Omega_{z,1} +  g_5 \Omega_{z,0}).
\end{align}

The functions $\phi_z^{(n)}$ in Eq.~\eqref{phiz-final} are
\begin{align}
\phi_z^{(0)} & \!= \!\frac{J}{\eta^4} \left(\!\frac{c_1^2}{2}\! \!- \frac{c_1 \eta^2}{6 v} \!-\! \frac{S^2_{\av}\eta^2}{3}\! -\!\frac{\eta^4}{3  v^2} \!\right) \!-\! \frac{c_1}{2\eta}\left(\frac{c_1^2}{\eta^4} \!-\! \frac{S^2_{\av}}{\eta^2} \right) l_1,
\\
\phi_z^{(1)} & = -\frac{J}{2 \eta^2}\left(c_1 + \eta L\right) + \frac{1}{2\eta^3}\left(c_1 - \eta^2 S^2_{\av}\right) l_1,
\\
\phi_z^{(2)} & = -J + \sqrt{S^2_{\av}} l_2 - \frac{c1}{\eta} l_1,
\\
\phi_z^{(3)} & = J  v - \eta l_1 + \frac{c_1}{\sqrt{S^2_{\av}}} l_2,
\\
\phi_z^{(4)} & = \frac{J}{2S^2_{\av}}  v\left(c_1+ v S^2_{\av}\right) - \frac{1}{2 (S^2_{\av})^{3/2}}\left(c_1^2 - \eta^2 S^2_{\av}\right) l_2,
\\
\phi_z^{(5)} & = -J  v \left(\frac{c_1^2}{2(S^2_{\av})^2} - \frac{c_1  v}{6S^2_{\av}} - \frac{ v^2}{3} -\frac{\eta^2}{3 S^2_{\av}} \right) \nn
\\
&+ \frac{c_1}{2 (S^2_{\av})^{5/2}} \left(c_1^2 - \eta^2 S^2_{\av}\right) l_2,
\end{align}
where we have defined
\begin{align}
l_1 &= \ln{\left( c_1 + J \eta + L \eta  \right)},
\\
l_2 &= \ln{\left( c_1 + J  \sqrt{S^2_{\av}}  v+  S^2_{\av} v  \right)},
\end{align}
In the above expressions we keep the roots $S^2_+$ and $S^2_-$ constant and equal to their initial value. The complexity of the roots' PN expansion makes its use prohibitive. Note that we do not expand the roots at all, but rather use their initial value as a form of partial resummation to increase the accuracy of our results. We find that this approximation does not affect our final result for the GW significantly.

%---------------------------------------------------------------------------
\section{Justification of the $\phi_z$ calculation}
\label{phizexpressions}

The precession-averaged $\langle\dot{\phi}_z\rangle\sub{\pr}$ given in Eqs.~\eqref{phizdotfinal}-\eqref{D4-def} is exact. In principle, we could calculate $D_2$ and $D_4$ as functions of $ v$, substitute them in Eq.~\eqref{phizdotfinal}, and carry out a PN expansion and integration to obtain $\precav{{\phi}_z}=\phi_{z,-1}$. 

Though this approach should work, in practice we run into $2$ considerable problems. First, the resulting $\phi_{z,-1}$ is ill-behaved in the small mass ratio limit, despite never having assumed comparable masses.  Second, $\phi_{z,-1}$ diverges when, at any point in the evolution of a precessing system, the total spin angular momentum is (anti)aligned with the orbital angular momentum. We stress that this does~\emph{not} mean that $\bm{S}$ is approximately (anti)aligned with $\bm{L}$ all the time; a brief moment of (anti)alignment suffices. 

Both issues are~\emph{not} caused by real physical divergences in $\Omega_z$. First, at no point did we assume comparable masses. The second issue is more subtle. It might be true that the denominator of $\Omega_z$ vanishes if $\bm{S}$ and $\bm{L}$ are (anti)aligned. However, the binary (and $\phi_z$) is well behaved at the moment of (anti)alignment since the numerator of $\Omega_z$ vanishes too, leading to a $0/0$ type situation{\footnote{We have verified that this is the case both analytically and numerically.}}. 

We argue that even though Eq.~\eqref{phizdotfinal} is well behaved in both the small mass ratio and the (anti)alignment between $\bm{S}$ and $\bm{L}$ limit, the same need not be true for its PN expansion. Consider the following function\footnote{The similarity between our toy function and J given in Eq.~\eqref{Jsol} is not accidental.}
\be
h(x;h_2,h_1,h_0) = \sqrt{\frac{h^2_2}{x^2}+\frac{h_1}{x} + h_0},
\ee
and its expansion around $x=0$
\be
h^{\sub{exp}}(x;h_2,h_1,h_0) = \frac{h_2}{x}+\frac{h_1}{2h_2}-\frac{h_1^2-4h_0h_4}{8h_2^3}x + {\cal{O}}(x^2).
\ee

Clearly, $h(x)$ is finite as $h_2\rightarrow0$. However, $h^{\sub{exp}}(x)$ is not, and the $h_2\rightarrow0$ limit is worse and worse as we keep more terms in the $x$ expansion.

This is exactly the situation we encounter with Eq.~\eqref{phizdotfinal} both in the small mass ratio limit, and in the approximate $\bm{S}$ and $\bm{L}$ (anti)alignment limit. Fixing the small mass ratio limit is straightforward: we identify the problem as originating from expanding the $J$ multiplying the entire right-hand side of Eq.~\eqref{phizdotfinal}, and factor it out. This is the reason behind the form Eq.~\eqref{int-phiz} has.

The second problem is more complicated. We can still identify the terms that, when expanded, cause the limit when $\bm{S}$ and $\bm{L}$ are (anti)aligned to be problematic. However, if we do not expand them, we can no longer perform the integral of Eq.~\eqref{int-phiz}. Using this fully expanded $\precav{\phi_z}$ causes $~5\%$ of the systems studied here to have faithfulnesses below threshold (see Sec.~\ref{sec:match}).

In light of this, we tried a number of alternative, approximate methods for calculating $\precav{{\phi}_z}$. We discovered that if we keep the terms $D_2$ and $D_4$ in Eqs.~\eqref{D2-def} and~\eqref{D4-def}, our results are greatly improved by about an order of magnitude: only $~0.8\%$ of the systems are below the faithfulness threshold (see Sec.~\ref{sec:match}). We examined a number of different definitions for $D_2$ and $D_4$, from using their initial value as given directly from Eqs.~\eqref{D2-def} and~\eqref{D4-def} to retaining different orders in a PN expansion, but evaluated at the initial time. We found out that these methods give comparable results, so we choose, for simplicity, to set $D_2$ and $D_4$ equal to their leading PN order:
\begin{align}
D_2 &\rightarrow \frac{c_p-\sqrt{c_p c_m}}{R_m \eta^2},\label{D2-value}
\\
D_4 &\rightarrow \frac{cp(c_p-\sqrt{c_p c_m})}{R_m^2 \eta^4} -\frac{\sqrt{c_pc_m}}{2R_m \eta^2}, \label{D4-value}
\end{align}
where $c_p, c_m, R_m$ are defined in Appendix~\ref{coeff-phiz}.

We expect this problem to be solved if we consistently PN expand both $(D_2,D_4)$ and the roots $S^2_+,S^2_-$ (see Appendix~\ref{coeff-phiz}). The complexity of the roots' expansion poses some serious problems in this calculation and we here opt for the approach described above and the partial resummation of the roots explained in Appendix~\ref{coeff-phiz}. This approach yields satisfactory results for the waveform precision required for aLIGO (see Fig.~\ref{fig:matches}), but can be improved if need be through expansions appropriate for these systems, like a small misalignment between $\bm{S}$ and $\bm{L}$ expansion.

%---------------------------------------------------------------------------
\section{Coefficients of $\zeta$}
\label{coeff-zeta}

The angle $\zeta$ that enters in the transformation to the waveform to the frame corotating with the precession of $\bm{L}$ can be calculated by solving 
\be
\dot{\zeta} = \dot{\phi_z} \cos{\theta_L}= \Omega_z \cos{\phi_L} \equiv \Omega_{\zeta}.
\ee

The solution to this equation can be obtained through MSA and it is very similar to the solution to Eq.~\eqref{phizdot}
\be
\zeta =\zeta_{-1}+  \zeta_{0}, 
\ee
where
\begin{align}
 \zeta_{-1} &= \int \precav{\Omega_{\zeta}}\!(t_{\rr}) \; dt_{\rr},\\
\zeta_{0} &= \int \left[ \Omega_{\zeta}(t_{\pr}, t_{\rr}) -  \precav{\Omega_{\zeta}}\!(t_{\rr})\right] \; dt_{\pr}.\label{zeta0-def}
\end{align}

Following the same steps as in Sec.~\ref{leading-msa} and for reasons explained in Appendix~\ref{phizexpressions} we find%
\be
\zeta_{-1} = \eta  v^{-3}\sum_{i=0}^{5}\langle \Omega_{\zeta}\rangle^{(n)}  v^{n} +\zeta^{0}_{-1},
\ee
where $\zeta^{0}_{-1}$ is a constant of integration and we have defined
\begin{align}
\langle \Omega_{\zeta}\rangle^{(0)}& = - g_0 \Omega_{\zeta,0},
\\
\langle \Omega_{\zeta}\rangle^{(1)} & = -\frac{3}{2}  g_0 \Omega_{\zeta,1},
\\
\langle \Omega_{\zeta}\rangle^{(2)} & = -3 ( g_0 \Omega_{\zeta,2} +  g_2 \Omega_{\zeta,0}),
\\
\langle \Omega_{\zeta}\rangle^{(3)}& = 3 ( g_0 \Omega_{\zeta,3}+  g_2 \Omega_{\zeta,1} +  g_3\Omega_{\zeta,0}),
\\
\langle \Omega_{\zeta}\rangle^{(4)} & = 3 ( g_0\Omega_{\zeta,4} +  g_2 \Omega_{\zeta,2} +  g_3 \Omega_{\zeta,1} +  g_4 \Omega_{\zeta,0}),
\\
\langle \Omega_{\zeta}\rangle^{(5)} & \!= \!\frac{3}{2} ( g_0\Omega_{\zeta,5}+  g_2\Omega_{\zeta,3} +  g_3 \Omega_{\zeta,2} +  g_4 \Omega_{\zeta,1} +  g_5\Omega_{\zeta,0}).
\end{align}
and
\begin{align}
\Omega_{\zeta,0} &= \Omega_{z,0},
\\
\Omega_{\zeta,1} &=\Omega_{z,1} +  \frac{c_1}{\eta^2}\Omega_{z,0},
\\
\Omega_{\zeta,2} &= \Omega_{z,2} + \frac{c_1}{\eta^2}\Omega_{z,1},
\\
\Omega_{\zeta,3}&= \Omega_{z,3} + \frac{c_1}{\eta^2}\Omega_{z,2} + g_d,
\\
\Omega_{\zeta,4} &= \Omega_{z,4} + \frac{c_1}{\eta^2}\Omega_{z,3} - g_d \xi - g_d h_d,
\\
\Omega_{\zeta,5} &= \Omega_{z,5} + \frac{c_1}{\eta^2}\Omega_{z,4} + g_d h_d \xi + g_d (h_d^2-f_d) ,
\end{align}
where the $\Omega_{z,i}$'s, $f_d$ and $h_d$ are given in Appendix~\ref{coeff-phiz} and
\be
g_d= \frac{3}{64} \frac{R_m^2}{\eta^3}\frac{c_1-\eta^2\xi}{\sqrt{c_pc_m}}.
\ee

The first correction to MSA is given by
\be
\zeta_{0} = \frac{A_{\theta_L}}{\dot{\psi}}(C_{\phi}+D_{\phi}) + 2 d_0\frac{ B_{\theta_L}}{\dot{\psi}}\left( \frac{C_{\phi}}{s_d-d_2}-\frac{D_{\phi}}{s_d+d_2}  \right),
\ee
where $C_{\phi}, D_{\phi}, d_0$ and $d_2$ are given in Appendix~\ref{coeff-phiz}, $\dot{\psi}$ is Eq.~\eqref{psi_rr}, and
\begin{align}
A_{\theta_L} & = \frac{J^2+L^2-S_+^2}{2 J L},
\\
B_{\theta_L} &=  \frac{S_+^2-S_-^2}{2 J L}.
\end{align}
%

%---------------------------------------------------------------------------
\section{Faithfulness requirement}
\label{f}

The agreement between two waveforms $h$, $\bar{h}$ with parameters $\vec{\lambda}$ is measured in terms of the faithfulness $F$
\begin{equation}
F(h,\bar{h}) = \frac{ ( h \vert \bar{h} )}{ \sqrt{ ( h \vert h ) (   \bar{h} \vert \bar{h} )}} \, .
\end{equation}
In the high SNR regime, a typical waveform sample from the posterior distribution function~\cite{Veitch:2014wba} is given by
\begin{equation}
\bar{h} = h + h_{,i} \Delta \lambda^i +\frac{1}{2} {h}_{,ij} \Delta \lambda^i \Delta \lambda^j + \dots
\end{equation}
where the $\Delta \vec{ \lambda} = \vec{\lambda} -  \vec{\lambda}_0$ are described by the multivariate normal distribution
\begin{equation}
p(\Delta \vec{ \lambda} ) = \sqrt{{\rm det}(\Gamma/2\pi)}e^{-\Gamma_{ij} \Delta \lambda^i  \Delta \lambda^j /2} \label{post}\, ,
\end{equation}
with $\Gamma_{ij}  =  (h_{,i} \vert  h_{,j})$ and $\vec{\lambda}_0$ are the true parameters.  Treating the $\Delta \lambda^i$ as small and expanding we get
\begin{equation}
F = 1 - \frac{1}{2} g_{ij} \Delta \lambda^i \Delta \lambda^j + \dots
\end{equation}
where
\begin{equation}
g_{ij}  = \frac{(h_i \vert h_j)}{(h \vert h)} - \frac{(h \vert h_{,i}) (h \vert h_{,j})}{(h \vert h)^2} \, .
\end{equation}
Using $E[ \Delta \lambda^i  \Delta \lambda^j] = C^{ij} \simeq \Gamma_{ij}^{-1}$, we find
\begin{equation}\label{EM}
E[F] \simeq 1 - \frac{(D-1)}{2\, {\rm SNR}^2} \, ,
\end{equation}
for the expectation value of the faithfulness, where $D$ is the dimension of $\vec{\lambda}$. The factor of $D$ comes from $C^{ij}\Gamma_{ij} \simeq \delta^i_i = D$ and
the factor of $-1$ from the $(h \vert h_{,i}) (h \vert h_{,j})/(h \vert h)^2$ removing the dependence on the overall amplitude of the waveform, thus reducing the dimensions count by one.

The expected value of the faithfulness in Eq.~(\ref{EM}) describes the impact of statistical errors. In deciding how accurate a waveform model needs to be, we should at a minimum demand
that the systematic errors from mismodeling are smaller than the statistical errors. If we wish to model spin-precessing binaries with $D=8$ intrinsic parameters for systems with SNRs up to 50
then the modeling unfaithfulness should be below $8/5000 = 0.0016$  (there is no $-1$ for just intrinsic parameters, the amplitude is extrinsic). For an SNR of $25$ we obtain the faithfulness requirement of $0.994$ that we used in Sec.~\ref{sec:match}.

To calculate the variance, it is easier to work with the unfaithfulness, $1-F$. The expectation of the square is given by
\begin{eqnarray}
E[(1-F)^2] &=& \frac{1}{4} g_{ij} g_{kl} E[ \Delta \lambda^i \Delta \lambda^j  \Delta \lambda^k \Delta \lambda^l]  \nonumber \\
&=& \frac{1}{4} g_{ij} g_{kl} \left( C^{ij} C^{kl} +  C^{ik} C^{jl} +  C^{il} C^{jk} \right)  \nonumber \\
&\simeq & \frac{3(D-1)^2}{4\, {\rm SNR}^4}
\end{eqnarray}
Thus
\begin{equation}
{\rm var}[1-F] = \frac{2(D-1)^2}{4\, {\rm SNR}^4}
\end{equation}
This shows that the average faithfulness is slightly less than 1-$\sigma$ from a perfect faithfulness ($\sigma/\sqrt{2}$ to be precise). This agrees with what we see when computing
the distribution of the match from MCMC waveform samples. The distribution is not Gaussian, and has a larger tail toward small values of the match.

An alternative derivation of the faithfulness requirement makes direct use of the posterior distribution function in the case of uniform priors
\begin{equation}
p(\vec{ \lambda} ) \sim e^{-\frac{(d-h|d-h)}{2}} \, ,
\end{equation}
where $d$ is the data. The peak of the posterior, evaluated at the best-fit parameters is
\begin{align}
p(\vec{ \lambda}_{bf} ) &\sim e^{-\frac{(d-h_{bf}|d-h_{bf} )}{2}} \sim e^{-\frac{(d|d)+(h_{bf}|h_{bf})-2(d-h_{bf})}{2}}\nn
\\
&\sim e^{-\frac{\text{SNR}^2+\text{SNR}^2-2 \text{SNR}^2\, \text{FF}}{2}}\sim e^{-\text{SNR}^2(1- \text{FF})},
\end{align}
where FF is the fitting factor, or the faithfulness maximized over all model parameters. The posterior on the true parameters is
\begin{align}
p(\vec{ \lambda}_0 ) &\sim e^{-\frac{(d-h_0|d-h_0 )}{2}} \sim e^{-\text{SNR}^2(1- \text{F})},
\end{align}

From Eq.~\eqref{post} we can calculate the value of the multidimensional posterior $1-\sigma$ away from the best-fit parameters
\begin{equation}
p(\vec{ \lambda}_{1-\sigma}) \sim e^{-\frac{ \Gamma_{ij}\Delta \lambda^i  \Delta \lambda^j }{2}}\sim e^{-\frac{ \Gamma_{ij}C^{ij} }{2}} \sim e^{-\frac{ D }{2}} \, ,
\end{equation}

Assuming that the model can fit the data perfectly for some parameters (an assumption that will lead to a conservative faithfulness threshold) we set FF=1 and requiring that the true parameters are less than $1-\sigma$ away from the best-fit ones we find
\begin{align}
1- \text{F}< \frac{ D }{2\text{SNR}^2} ,
\end{align}
where $D$ is the number of parameters whose measurability is affected by the model inaccuracy. For spin-precessing models with $8$ intrinsic parameters, $D=8$.

This derivation translates the results of~\cite{Lindblom:2008cm} that were written in terms of requirements on the GW amplitude and phase to requirement on the faithfulness.

%%%%%%%%%%%%%%%%%%%%%%%%%%%%%
%%%%%%%%%%%%%%%%%%%%%%%%%%%%%
\bibliography{review}
\end{document}